\documentclass[aps,prd,showpacs,amssymb,amsmath,amsfonts,superscriptaddress,
twocolumn,floatfix,preprintnumbers,altaffilletter]{revtex4}
\newif\ifpdf\ifx\pdfoutput\undefined\pdffalse\else\pdfoutput=1\pdftrue\fi
       \newcommand{\pdfgraphics}{\ifpdf\DeclareGraphicsExtensions{.pdf,.jpg}\else\fi}
\usepackage{graphicx}
\usepackage{hyperref}
\usepackage{amssymb}
\usepackage{amsmath}
\usepackage{longtable}
\usepackage{rotating}
\usepackage{color}
\usepackage{epsfig}
\usepackage{epsf}
\usepackage{fancyhdr}

\def\th{\textrm{\mbox{\tiny{th}}}}
\def\Tobs{T_{\textrm{\mbox{\tiny{obs}}}}}
\def\Tcoh{T_{\textrm{\mbox{\tiny{coh}}}}}

\def\min{\textrm{\mbox{\tiny{min}}}}
\def\max{\textrm{\mbox{\tiny{max}}}}

\def\inject{\textrm{\mbox{\tiny{inject}}}}
\def\Exp{\textrm{\mbox{\tiny{exp}}}}
\def\band{\textrm{\mbox{\tiny{band}}}}

\newcommand{\ee}[1]{\!\times\!10^{#1}}

\newcommand{\F}{\mathcal{F}}
\def\bea{\begin{eqnarray}}
\def\eea{\end{eqnarray}}
\def\beq{\begin{equation}}
\def\eeq{\end{equation}}

\def\be{\begin{equation}}
\def\ee{\end{equation}}

\newcommand{\ba}{\begin{eqnarray}}
\newcommand{\ea}{\end{eqnarray}}

\newcommand{\bml}{\begin{mathletters}}
\newcommand{\eml}{\end{mathletters}}

\begin{document}
\pdfgraphics
\pagestyle{fancy}
%\chead[{\color{red} \large Circulation restricted to LIGO-I members}]
%      {{\color{red} \large Circulation restricted to LIGO-I members}}
\rhead[]{}
\lhead[]{}
\title{ %{{\color{red} \large Circulation restricted to LIGO-I members}}\\
 First all-sky upper limits from LIGO on the strength of
  periodic gravitational waves using the Hough transform}

%\title{ {{\color{red} \large Circulation restricted to LIGO-I members}}\\
%All-sky search for continuous gravitational wave signals in LIGO data using
%  the Hough transform \\ or\\ First upper limits from LIGO on the strength of
%  periodic gravitational waves using the Hough transform.}

%********************* cut here mark*******************
%NOTE ADD altafillletter to \documentclass if lettered affiliation footnotes are desired
% EXAMPLE:
%
%\documentclass[prd,superscriptaddress,showpacs,amssymb,amsmath,amsfonts,aps,altaffilletter]{revtex4}
%
%%%%%%%%%%%% Institutes Number and definitions %%%%%%%%%%%
%
% This list was generated on June 20 2005 A4 DCC LIGO - T040086-14-Z
\newcommand*{\AG}{Albert-Einstein-Institut, Max-Planck-Institut f\"ur Gravitationsphysik, D-14476 Golm, Germany}
\affiliation{\AG}
\newcommand*{\AH}{Albert-Einstein-Institut, Max-Planck-Institut f\"ur Gravitationsphysik, D-30167 Hannover, Germany}
\affiliation{\AH}
\newcommand*{\AN}{Australian National University, Canberra, 0200, Australia}
\affiliation{\AN}
\newcommand*{\CH}{California Institute of Technology, Pasadena, CA  91125, USA}
\affiliation{\CH}
\newcommand*{\DO}{California State University Dominguez Hills, Carson, CA  90747, USA}
\affiliation{\DO}
\newcommand*{\CA}{Caltech-CaRT, Pasadena, CA  91125, USA}
\affiliation{\CA}
\newcommand*{\CU}{Cardiff University, Cardiff, CF2 3YB, United Kingdom}
\affiliation{\CU}
\newcommand*{\CL}{Carleton College, Northfield, MN  55057, USA}
\affiliation{\CL}
\newcommand*{\CO}{Columbia University, New York, NY  10027, USA}
\affiliation{\CO}
\newcommand*{\HC}{Hobart and William Smith Colleges, Geneva, NY  14456, USA}
\affiliation{\HC}
\newcommand*{\IU}{Inter-University Centre for Astronomy  and Astrophysics, Pune - 411007, India}
\affiliation{\IU}
\newcommand*{\CT}{LIGO - California Institute of Technology, Pasadena, CA  91125, USA}
\affiliation{\CT}
\newcommand*{\LM}{LIGO - Massachusetts Institute of Technology, Cambridge, MA 02139, USA}
\affiliation{\LM}
\newcommand*{\LO}{LIGO Hanford Observatory, Richland, WA  99352, USA}
\affiliation{\LO}
\newcommand*{\LV}{LIGO Livingston Observatory, Livingston, LA  70754, USA}
\affiliation{\LV}
\newcommand*{\LU}{Louisiana State University, Baton Rouge, LA  70803, USA}
\affiliation{\LU}
\newcommand*{\LE}{Louisiana Tech University, Ruston, LA  71272, USA}
\affiliation{\LE}
\newcommand*{\LL}{Loyola University, New Orleans, LA 70118, USA}
\affiliation{\LL}
\newcommand*{\MP}{Max Planck Institut f\"ur Quantenoptik, D-85748, Garching, Germany}
\affiliation{\MP}
\newcommand*{\MS}{Moscow State University, Moscow, 119992, Russia}
\affiliation{\MS}
\newcommand*{\ND}{NASA/Goddard Space Flight Center, Greenbelt, MD  20771, USA}
\affiliation{\ND}
\newcommand*{\NA}{National Astronomical Observatory of Japan, Tokyo  181-8588, Japan}
\affiliation{\NA}
\newcommand*{\NO}{Northwestern University, Evanston, IL  60208, USA}
\affiliation{\NO}
\newcommand*{\SC}{Salish Kootenai College, Pablo, MT  59855, USA}
\affiliation{\SC}
\newcommand*{\SE}{Southeastern Louisiana University, Hammond, LA  70402, USA}
\affiliation{\SE}
\newcommand*{\SA}{Stanford University, Stanford, CA  94305, USA}
\affiliation{\SA}
\newcommand*{\SR}{Syracuse University, Syracuse, NY  13244, USA}
\affiliation{\SR}
\newcommand*{\PU}{The Pennsylvania State University, University Park, PA  16802, USA}
\affiliation{\PU}
\newcommand*{\TC}{The University of Texas at Brownsville and Texas Southmost College, Brownsville, TX  78520, USA}
\affiliation{\TC}
\newcommand*{\TR}{Trinity University, San Antonio, TX  78212, USA}
\affiliation{\TR}
\newcommand*{\HU}{Universit{\"a}t Hannover, D-30167 Hannover, Germany}
\affiliation{\HU}
\newcommand*{\BB}{Universitat de les Illes Balears, E-07122 Palma de Mallorca, Spain}
\affiliation{\BB}
\newcommand*{\BR}{University of Birmingham, Birmingham, B15 2TT, United Kingdom}
\affiliation{\BR}
\newcommand*{\FA}{University of Florida, Gainesville, FL  32611, USA}
\affiliation{\FA}
\newcommand*{\GU}{University of Glasgow, Glasgow, G12 8QQ, United Kingdom}
\affiliation{\GU}
\newcommand*{\MU}{University of Michigan, Ann Arbor, MI  48109, USA}
\affiliation{\MU}
\newcommand*{\OU}{University of Oregon, Eugene, OR  97403, USA}
\affiliation{\OU}
\newcommand*{\RO}{University of Rochester, Rochester, NY  14627, USA}
\affiliation{\RO}
\newcommand*{\UW}{University of Wisconsin-Milwaukee, Milwaukee, WI  53201, USA}
\affiliation{\UW}
\newcommand*{\VC}{Vassar College, Poughkeepsie, NY 12604}
\affiliation{\VC}
\newcommand*{\WU}{Washington State University, Pullman, WA 99164, USA}
\affiliation{\WU}

\author{B.~Abbott}    \affiliation{\CT}
\author{R.~Abbott}    \affiliation{\LV}
\author{R.~Adhikari}    \affiliation{\CT}
\author{A.~Ageev}    \affiliation{\MS}  \affiliation{\SR}
\author{J.~Agresti}    \affiliation{\CT}
\author{B.~Allen}    \affiliation{\UW}
\author{J.~Allen}    \affiliation{\LM}
\author{R.~Amin}    \affiliation{\LU}
\author{S.~B.~Anderson}    \affiliation{\CT}
\author{W.~G.~Anderson}    \affiliation{\TC}
% ...

 \collaboration{The LIGO Scientific Collaboration, http://www.ligo.org}
 \noaffiliation
%
%
%********************* cut here mark*******************

%\author{The LIGO Scientific Collaboration \\ (Author list TBD)}
%
\date{\today}
%\date{$Revision: 1.219 $,  $Date: 2005/08/16 09:07:26 $}
%$Id: S2HoughPaper.tex,v 1.219 2005/08/16 09:07:26 sintes Exp $}
%
\begin{abstract}
  We perform a wide parameter space search for continuous gravitational
  waves over the whole sky and over a large range of values of the
  frequency and  the first spin-down parameter.  Our search method is
  based on the Hough transform, which is a semi-coherent,
  computationally efficient, and robust pattern recognition technique.
  We apply this technique to data from the second science run of the
  LIGO detectors and our final results are all-sky upper limits
  on the strength of gravitational waves emitted by unknown isolated
  spinning neutron stars on a set of narrow frequency bands in the 
  range $200$-$400\,$Hz.  The best upper limit on the gravitational wave
  strain amplitude that we obtain in this frequency range is
  $4.43\times 10^{-23}$.    
\end{abstract}
\preprint{LIGO-P050013-03-R}
\maketitle
%
%%%%%%%%%%%%%%%%%%%%%%%%%%%%%%%%%%%%%%%%%%%%%%%%%%%%%%%%%%%%%%%%%%%%%%%%%%%%%%%%%%%%%%%%
\section{Introduction}
\label{sec:intro}

Continuous gravitational signals emitted by rotating neutron
stars are promising sources for interferometric gravitational wave detectors
such as GEO\,600 \cite{GEO1,GEO2}, the Laser Interferometer 
Gravitational Wave Observatory (LIGO) \cite{ligo1, ligo2}, TAMA\,300
\cite{tama95} and VIRGO \cite{virgo97}. 
There are several physical mechanisms which might cause a
neutron star to emit periodic gravitational waves.
The main possibilities considered in the literature are 
(i)~non-axisymmetric distortions of the solid part of the star
\cite{Bildsten:1998ey, UCB00, Cutler02, Owen:2005fn}, 
(ii)~unstable $r$-modes in the fluid \cite{Owen:1998xg, Bildsten:1998ey,
Andersson:1998qs}, and
(iii)~free precession (or `wobble') \cite{DIJones,vdb04}.
The detectability of a signal depends on the detector sensitivity, the
intrinsic emission strength, the source distance and its orientation. 
If the source is not known, the detectability also depends on the available
computational resources.   
For some search methods the detectability of a signal also depends on the
source model used, but an all-sky wide-band search such as detailed here can
detect any of the sources described above.

Previous searches for gravitational waves from rotating neutron stars have
been of two kinds.  The first is a search targeting 
pulsars whose parameters are known through radio observations.  These
searches typically use matched filtering techniques and are not very
computationally expensive. 
Examples of such searches are \cite{S1-CW} and \cite{cw-prl}
which targeted known radio pulsars, at twice the pulsar frequency,
 using data from the first and second
science runs of the GEO\,600 and LIGO detectors \cite{S1_exp}.
No signals were detected and the end results were upper limits on the
strength of the gravitational waves emitted by these pulsars and
therefore on their ellipticity. 

The second kind of search looks for as yet undiscovered rotating
neutron stars.  An example of such a search is \cite{astone}
in which a two-day long data stretch from the Explorer bar detector is
used to perform an all sky search in a narrow frequency band around
the resonant frequency of the detector. Another example is
\cite{S2-FDS} which uses data from the LIGO detectors to perform
an all sky search in a wide frequency band using  10 hours of
data. The same paper also describes a search for a gravitational
  wave signal from the compact companion to Sco X-1 in a large
  {\it orbital} parameter space using 6 hours of data. 
  The key issue in these wide parameter space searches is that a
fully coherent all-sky search over a large frequency band using a
significant amount of data is computationally limited.  This is because 
looking for
weak continuous wave signals requires long observation times to build
up sufficient signal to noise ratio;
% and to claim a detection with some
%degree of confidence; 
the amplitude signal-to-noise ratio increases as
the square root of the observation time.  On the other hand, the
number of templates that must be considered, and therefore the
computational requirements, scale much faster than linearly with the
observation time.  We therefore need methods which
are sub-optimal but computationally less expensive 
\cite{BCCS,BC00,cgk,hough04,pss01}. 
Such methods
typically involve semi-coherent combinations of the signal power in
short stretches of data, and the Hough transform is an example of such
a method.  

The Hough transform is a pattern recognition algorithm which was
originally invented to analyze bubble chamber pictures from CERN
\cite{hough1}. It was later patented by IBM \cite{hough2}, and it has
found many applications in the analysis of digital images \cite{ik}.
A detailed discussion of the Hough transform as applied to the search
for continuous gravitational waves can be found in \cite{hough04,
  hough05}. In this paper, we apply this technique to perform an
all-sky search for isolated spinning neutron stars using two months of
data collected in early 2003 from the second 
science run of the LIGO detectors (henceforth denoted as the S2 run).
The main results of this paper are  all-sky upper limits on a set
of narrow frequency bands within the range $200$-$400\,$Hz and
including one spin-down parameter. 

Given the detector strain sensitivities during the S2 run and its duration,
it is unlikely to discover any neutron star, as described in the following
Sec.~\ref{subsec:targets}. 
 For this reason we focus here on setting upper limits.
Substantial improvements in the detector noise have been achieved
since the S2 observations. A third science run (S3) took place at the end
of 2003 and a fourth science run (S4) at the beginning of 2005.
In these later runs  LIGO instruments collected data of improved sensitivity,
but still less sensitive than the instruments' design goal.
Several searches for various types of gravitational wave signals have
been completed or are underway using data from the S2 and S3 runs 
\cite{cw-prl,S2-FDS,S2-GRB,S2-burst,S2-NS,S2-BH,S2-tama,ligo-BHM,S3-sto}.
We expect that the method presented here, applied as part of a hierarchical 
scheme and used on a much larger data set, 
will eventually enable the direct detection of periodic gravitational
waves.  

%discusses

This paper is organized as follows:
Sec.~\ref{sec:ligos2} describes the second science run of the LIGO
detectors;
Sec.~\ref{subsec:targets} summarizes the current understanding of the
astrophysical targets;
%briefly discusses the astrophysical targets.
Sec.~\ref{subsec:waveform} reviews the waveform from an isolated spinning
neutron star;
Sec.~\ref{subsec:houghbasics} presents the general idea of 
 our search method, the Hough transform, and summarizes its statistical
 properties;
Sec.~\ref{sec:searchresults} describes its implementation and results
on short Fourier transformed data.
The upper limits are given in Sec.~\ref{sec:ul}.  
Sec.~\ref{sec:hardwareinj} presents a validation of
our search method using hardware injected signals, and finally 
Sec.~\ref{sec:conc} concludes with a summary of our results 
and suggestions for further work.

%%%%%%%%%%%%%%%%%%%%%%%%%%%%%%%%%%%%%%%%%%%%%%%%%%%%%%%%%%%%%%%%%%%%%%%%%%%%%%%%%%%%%%%%%%%
\section{The second science run}
\label{sec:ligos2}

%The behaviour of the detectors during S2.  What the Hough transform
%can hope to see with this data.  Spectral disturbances present in the
%data.  \marginnote{help}

%\input s2data.tex

The LIGO detector network consists of a 4 km interferometer in
Livingston, Louisiana (L1), and two interferometers in Hanford,
Washington, one 4 km and the other 2 km (H1 and H2).
Each detector is a power-recycled Michelson interferometer with
long Fabry-Perot cavities in each of its orthogonal arms.
%able to measure directly the gravitational wave strain amplitude.
These interferometers are sensitive to quadrupolar oscillations in
the space-time metric due to a passing gravitational wave, measuring
directly the gravitational wave strain amplitude.

The data analyzed in this paper were produced during LIGO's 59 day
second science run. This run started on February 14 and ended
April 14, 2003. Although the GEO detector was not operating at the
time, all three LIGO detectors were functioning at a significantly
better sensitivity than during S1, the first science
run \cite{S1_exp},
%All three LIGO detectors 
and had displacement spectral amplitudes near $10^{-18}$
m-Hz$^{-1/2}$ between $200\,$Hz and $400\,$Hz.  The strain sensitivities in 
this science run were within an order of magnitude of the design
sensitivity for the LIGO detectors.
For a description of the detector configurations 
for S2 we refer the
reader to \cite{S2-GRB} Sec.~IV and \cite{S2-burst} Sec.~II.

%The gravitational wave strain signal is derived from the error signal
%of the feedback loop used to control the differential length of the
%interferometer arms. To calibrate the error signal, the response to a
%known differential arm strain is measured, and compensated for.

The reconstruction of the strain signal from the error signal
of the feedback loop, used to control the differential length of the
interferometer arms, is referred to as the calibration.
Changes in the calibration were tracked by injecting continuous, 
fixed-amplitude
sinusoidal excitations into the end test mass control systems, and monitoring
the amplitude of these signals at the measurement error point. 
Calibration uncertainties at the three LIGO detectors during S2 
were estimated to be smaller than $11\%$ 
\cite{S2caldoc}.

The data were acquired and digitized at a rate of $16384\,$Hz.
The duty cycle for the interferometers, defined as the fraction of the total
run time when the interferometer was locked ({\it i.e.}, 
all interferometer control
servos operating in their linear regime)
and in its low noise configuration,
%Data acquisition was periodically interrupted by disturbances
%such as seismic transients or poorly conditioned servos,
%reducing the net running time of the interferometers.
%The resultant duty cycle for the interferometers
were similar to those of the previous science run, 
approximately 37$\%$ for L1, 74$\%$ for
H1 and 58$\%$ for H2. The longest continuous locked stretch for any
interferometer during S2 was 66 hours for H1.  The smaller 
duty cycle for L1 was due to anthropogenic diurnal low-frequency
seismic noise which prevented operations during the day on weekdays.
Recently installed active feedback seismic isolation has successfully
addressed this problem.

\begin{figure}
  \begin{center}
  \includegraphics[height=6.5cm]{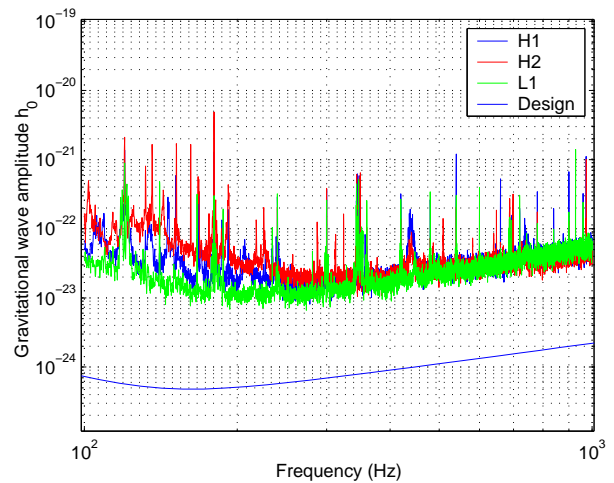}
  \caption{Characteristic amplitude detectable from a known generic 
  source with a 1\% false alarm rate and 10\% false dismissal
  rate, as given by Eq.~(\ref{eq:sensitivity}).
   All curves use typical sensitivities of the three LIGO
  detectors during S2 and observation times corresponding to the
  up-time of the detectors during S2.  The thin line is the expected
  characteristic amplitude for the same false alarm and false
  dismissal rates, %and $\Tcoh = 1800\,$s, 
  but using the initial LIGO design goal for the $4\,$km instruments 
  and an effective observation time of $1\,$yr.} 
  \label{fig:S2sensitivity} 
  \end{center}
\end{figure}
Fig.~\ref{fig:S2sensitivity} shows the expected sensitivity for the 
Hough search by the three LIGO detectors during S2. Those 
$h_0$ values correspond to the amplitudes detectable  from a
generic continuous gravitational wave source,  if we were performing  
a targeted search, with a 1\%  false alarm rate and 10\%  
false dismissal rate, as given by Eq.~(\ref{eq:sensitivity}).
%
%\be
%\label{eq:sensitivity1-10}
%h_0 =  \frac{8.54}{N^{1/4}}\sqrt\frac{S_n}{\Tcoh} \, 
%\ee
%(see Sec.~\ref{subsec:houghbasics} for details).
The differences among the three interferometers reflect differences in
the operating parameters, hardware implementation of the three
instruments, and the duty cycles.  
Fig.~\ref{fig:S2sensitivity} also shows the expected sensitivity (at
the same false alarm and false dismissal rates) for initial LIGO
$4\,$km interferometers running at design sensitivity
assuming an observation time of 1 year.
%, and assuming that we have 1 year's worth of data.  
%If we compare Eq.~(\ref{eq:sensitivity1-10}) with Eq.~(2.2) in \cite{S1-CW},
%it turns out that a Hough search would be less sensitive than a full directed
%two months coherent search by only a factor of 5. 
%Because of the large parameter search we perform here, it would be
%more meaningful to consider a lower false alarm rate. 
%The exact value would depend on the number of candidates
%we would like to select and the number of templates analyzed. 
%At a false alarm rate say of  $10^{-10}$, the sensitivity for a targeted 
%search would get worse by a factor $1.5$  as compared 
%to (\ref{eq:sensitivity1-10}).  
These false alarm and false dismissal values are chosen in agreement
with \cite{S1-CW} and  \cite{S2-FDS} only for comparison
purposes. Because of the large parameter search we perform here, it
would be more meaningful to consider a lower false alarm rate, say of
$10^{-10}$ and then the sensitivity for a targeted search would get
worse by a factor $1.5$.  The search described in this paper is not
targeted and this degrades the sensitivity even further. This will be
discussed in Sec.~\ref{sec:ul}.

At the end of the S2 run, two fake artificial pulsar signals  
were injected for a 12 hour period into all three LIGO interferometers. 
These hardware injections were done by modulating the mirror positions
via the actuation control signals. %\cite{uta, uta04}.  
These injections
were designed to give an end-to-end validations of the search
pipelines starting from as far up the observing chain as possible.
See Sec.~\ref{sec:hardwareinj} for details. 

\section{Astrophysical targets}
\label{subsec:targets}

The target population of this search consists of isolated rotating
neutron stars that are not observed in electromagnetic waves.
Current models of stellar evolution suggest that our Galaxy contains of
order $10^9$ neutron stars and that of order $10^5$ are active
pulsars \cite{lorimer}.
Up to now, only of order $10^3$ objects have been identified as
neutron stars, either by observation as pulsars, or through their
x-ray emission, of which about 90\% are isolated \cite{atnf,Haberl:2003cg,Pavlov:2003eg,KaspiEtAl2004}.
Most neutron stars will remain unobserved electromagnetically for many
reasons such as the non-pulsed emission being faint or the pulses being
emitted in a beam which does not sweep across the Earth.  Therefore,
there are many more neutron stars in the target population than have
already been observed.

Although there is great uncertainty in the physics of the emission
mechanism and the strength of an individual source, we can argue for a
robust upper limit on the strength of the strongest source in the galactic
population that is almost independent of individual source physics. The
details of the argument and an overview of emission mechanisms can be
found in a forthcoming paper~\cite{S2-FDS}. Here we do not repeat the details
but merely summarize the result. For an upper limit we make optimistic
assumptions---that neutron stars are born rapidly rotating and spinning
down due to gravitational waves, and that they are distributed uniformly
throughout the galactic disc---and the plausible assumption that the
overall galactic birthrate $1/\tau_b$ is steady. By converting these
assumptions to a distribution of neutron stars with respect to
gravitational wave strain and frequency, 
we find there is a $50\%$ chance that the strongest signal between 
frequencies $f_{\min}$ and $f_{\max}$ has an amplitude of at least
\begin{equation}
\label{cutler1}
h_0 \approx 4\times10^{-24} \left[ \left( 30\mbox{ yr} \over \tau_b
\right) \ln{f_{\max} \over f_{\min}} \right]^{1/2}.
\end{equation}
Of course, with less optimistic assumptions this value would be smaller.

Comparing Eq.~(\ref{cutler1}) to Fig.~\ref{fig:S2sensitivity}, a search of
S2 data is not expected to result in a discovery.
However, it is still possible that the closest neutron star is closer than
the typical distance expected from a random distribution of supernovae (for
example due to recent star formation in the Gould belt as considered in
Ref.~\cite{palomba}).
It is also possible that a ``blind'' search of this sort may discover some
previously unknown class of compact objects not born in supernovae.
More importantly, future searches for previously undiscovered rotating
neutron stars using the methods presented here will be much more sensitive.
The goal of initial LIGO is to take a year of data at design sensitivity,
which means a factor 10 decrease in the amplitude strain noise
relative to S2, and a factor
%$10^2$ 
$10$ increase in the length of the data set.
These combine to reduce $h_0$ %by a factor 20--30 
%I guest a factor 3-50 depending if coherent or Hough search is used
to somewhat below the value
in Eq.~(\ref{cutler1}), and thus initial LIGO at full sensitivity will have
some chance of observing a periodic signal.

%%%%%%%%%%%%%%%%%%%%%%%%%%%%%%%%%%%%%%%%%%%%%%%%%%%%%%%%%%%%%%%%%%%%%%%%%%%%%%%%%%%%
\section{The expected waveform}
\label{subsec:waveform}

In order to describe the expected signal waveform we will use the same 
notation as \cite{S1-CW}. We will briefly summarize it in the next paragraphs 
for convenience. 
The form of the gravitational wave emitted by an isolated spinning neutron star,
 as seen by a
gravitational wave detector, is 
\begin{equation}
h(t) = F_+(t,\psi)h_+(t) + F_\times(t,\psi)h_\times(t) \,,\label{eq:detoutput}
\end{equation}
where $t$ is time in the detector frame, $\psi$ is the polarization
angle of the wave, and $F_{+,\times}$ are the detector antenna pattern
functions for the two polarizations.  If
we assume the emission mechanism is due to deviations of the pulsar's
shape from perfect axial symmetry, then the gravitational waves are
emitted  at a frequency which is exactly twice the rotational rate $f_r$. 
 Under this assumption, the waveforms for the two
polarizations $h_{+,\times}$ are given by:  
\begin{eqnarray}
h_+ &=& h_0 \frac{1+ \cos^2\iota}{2}\cos\Phi(t)\,, \label{eq:waveform1}\\
h_\times &=& h_0 \cos\iota\sin\Phi(t)\,, \label{eq:waveform2}
\end{eqnarray}
where $\iota$ is the angle between the neutron star's spin axis and the
direction of propagation of the waves, and $h_0$ is the amplitude:
\begin{equation} \label{eq:h0} h_0 = \frac{16\pi^2G}{c^4}\frac{I_{zz}\epsilon
f_r^2}{d}\, , \end{equation}
where $G$ is Newton's gravitational constant, $c$ the speed of light,
$I_{zz}$ is the principal moment with the $z$-axis being
its spin axis,  $\epsilon:= (I_{xx}-I_{yy})/I_{zz}$ 
is the equatorial ellipticity of the star, and
$d$ is the distance to the star.

The phase $\Phi(t)$ takes its simplest form in the Solar System Barycenter
(SSB) frame where it can be expanded in a Taylor series up to second order:
\begin{equation} \label{eq:phasemodel}
\Phi(t) = \Phi_0 + 2\pi\left( f_0(T - T_0) +
\frac{1}{2}\dot{f}(T-T_0)^2 \right)\,.
\end{equation}
Here $T$ is time in the SSB frame and $T_0$ is a fiducial start time.
The phase $\Phi_0$, frequency $f_0$ and  spin-down parameter 
$\dot{f}$ are defined
at this fiducial start time.  In this paper, we include only one
spin-down parameter in our search; as we shall see later in
Sec.~\ref{subsec:parameters}, our frequency resolution is too coarse for the
higher spin-down parameters to have any significant effect on the
frequency evolution of the signal (for the spin-down ages we
consider). 

Modulo relativistic effects which are unimportant for this search, the
relation between the time of arrival $T$ of the wave in the SSB frame
and in the detector frame $t$ is   
\begin{equation}
T = t + \frac{\mathbf{n\cdot r}}{c}\,,
\end{equation}
where $\mathbf{n}$ is the unit vector from the detector to the neutron star,
and $\mathbf{r}$ is the detector position in the SSB frame.  

The instantaneous frequency $f(t)$ of the wave as observed by the
detector is given, to a very good approximation, by the familiar 
non-relativistic Doppler formula: 
\begin{equation}\label{eq:master}
f(t) - \hat{f}(t) = \hat{f}(t)\frac{ {\bf v} (t)\cdot\bf{n}}{c} \,,
\end{equation}
where $\hat{f}(t)$ is the instantaneous signal frequency in the SSB frame at time $t$: 
\begin{equation} \label{eq:masterspndn}
\hat{f}(t) = f_0 + \dot{f}\left(t -t_0 + 
\frac{\Delta\mathbf{r}(t)\cdot\mathbf{n}}{c}\right) \,,
\end{equation}
where $t_0$ is the fiducial detector time at the start 
of the observation and $\Delta\mathbf{r}(t) = \mathbf{r}(t) -
\mathbf{r}(t_0)$.  It is easy to see that the
$\Delta\mathbf{r}\cdot\mathbf{n}/c$ term can safely be ignored so that, to
an excellent approximation
\begin{equation} \label{eq:fhat}
\hat{f}(t) = f_0 + \dot{f}\left(t -t_0 \right)\,. 
\end{equation}
%
%Eq.s~(\ref{eq:master}) and (\ref{eq:fhat}) describe the
%time-frequency pattern produced by a signal, and we employ the Hough
%transform to look for this particular pattern.

%%%%%%%%%%%%%%%%%%%%%%%%%%%%%%%%%%%%%%%%%%%%%%%%%%%%%%%%%%%%%%%%%%%%%%%%%%%%%%%%%%%%%
\section{The Hough transform}
\label{subsec:houghbasics}

In this paper, we use the Hough transform to find the pattern produced by the 
Doppler shift (\ref{eq:master}) and the spin-down (\ref{eq:fhat}) of 
a gravitational wave signal in the time-frequency plane of our data. 
This pattern is independent of the source model used and therefore
of the emission mechanisms. We only assume that the gravitational wave
signal is emitted by an isolated spinning neutron star.

%The starting point for the Hough transform are $N$ short stretches of Fourier transformed
%data; each short stretch will be called an SFT (Short Fourier Transform). 
	%The input to our search pipeline is a
%The starting point for the Hough transform is a

The starting point for our search  is a
set of data segments, each corresponding to a time interval $\Tcoh$.
Each of these data segments is Fourier transformed to
produce a set of $N$ short time-baseline Fourier transforms (SFTs).
%We choose $\Tcoh = 30\,$min; see Sec.~\ref{subsec:sftdata} for the
%rationale behind this choice.   
%The duration $\Tcoh$ is chosen in such a way that the signal from a source
%would be  contained in a single frequency bin.
From this set of SFTs, calculating the periodograms (the square modulus of the
Fourier transform) and selecting 
frequency bins (peaks) above a certain threshold, 
we obtain a time-frequency map 
of our data. %, the so-called peak-grams. 
In the absence of a signal the peaks in the time-frequency plane %peak-grams
are distributed in a random way; if signal is present, with high enough signal
to noise ratio, some of these peaks will be distributed along the trajectory
of the received frequency of the signal.

The Hough transform maps points of the time-frequency plane into 
the space of the source parameters $(f_0,\dot{f},\mathbf{n})$.
The result of the Hough transform is a histogram, 
{\it i.e.}, a collection of integer numbers,
 each representing the detection statistic for each point in 
 parameter space. We shall refer to these integers as the number count.
The number counts are computed in the following way:
For each selected bin in the SFTs, we find which points in parameter 
 space are consistent with it, according to Eq.~(\ref{eq:master}), and
the number count in all such points is increased
by unity. This is repeated for all the selected  bins in all the
SFTs to obtain the final histogram.
 
%The parameters which determine this pattern
%are $(f_0,\dot{f},\mathbf{n})$. 

%This space is covered by a discrete
%grid whose resolution was described earlier.   
%Using the standard 
%language of matched filtering (even though this is a semi-coherent search 
%method) the patterns are the template bank for our search corresponding
%to  the grid in parameter space. 
%The set of patterns is the template bank for our search corresponding
%to  a grid in parameter space. 
%The result of the Hough transform is a histogram, 
%i.e. a collection of integer numbers,
% each representing the detection statistic for each point in 
% parameter space. We shall refer to these integers as the number count.

%How are these number counts computed? First, SFT frequency
%bins are selected based on a threshold on the power. 
%If the power in a bin exceeds the threshold value then that frequency bin 
%for that SFT is marked, else it is discarded. This yields, for every SFT, 
%a set of indexes, which represent the frequency bins that have been marked.
%	 %$i_{\alpha k}$, where $\alpha$ is the SFT order number and $k$ the frequency index. 
% For each selected  bin, we then find which points in parameter 
% space are consistent with it. % according to Eq.~(\ref{eq:master}).
%The chief virtue of the Hough transform is computational speed.
%This method 
%relies on marking all the possible templates consistent with
%a frequency bin in
%a particular SFT.  

To illustrate this, let us assume the source parameters are only the 
coordinates of the source in the sky, and this source is emitting 
a signal at a frequency $f_0$. Moreover we assume that at a given time $t$ 
a peak at frequency $f$ has been selected in the corresponding SFT.
The Hough transform maps this peak into the loci of points, on the celestial
sphere, where a source emitting  a signal with frequency $f_0$ could be 
located in order in order to produce at the detector a peak at $f$. By repeating this
for all the selected peaks in our data we will obtain the final Hough map.
If the peaks in the time-frequency plane were due only to signal, all 
the corresponding loci would intersect in a region of the Hough map 
identifying the source position.

% This is to be contrasted with another well
%known semi-coherent method, the stack-slide search \cite{BC00}, in 
%which one steps
%through the parameter space point by point and adds the power from the
%relevant frequency bin of each SFTs to obtain the final statistic, the
%summed power at each point of the parameter space grid.  The Hough transform
%avoids stepping through the parameter space point-by-point by throwing
%away the information on how much each observation contributes to the
%final statistic at any given template.
%; each observation can contribute
%either 0 or 1 to the final number count.  

An advantage of the Hough transform is that a large region in parameter 
space can be analyzed in a single pass. 
By dropping the amplitude information of the selected peaks,
 the Hough search is expected to be computationally
efficient, but at the cost of being somewhat less sensitive than others
semi-coherent methods, {\it e.g.}, the stack-slide search \cite{BC00}.  
On the other hand,  
discarding this extra information makes the Hough transform more
robust against transient spectral disturbances because no matter how
large a spectral disturbance is in a single SFT, it will contribute at
the most +1 to the number count. This is not surprising since the optimal 
statistic for the detection of weak signals in the presence of a Gaussian
 background with large non Gaussian outliers is effectively cut off above 
 some value \cite{C99,APS02}. This is, in practice, what the Hough
  transform does to large spectral outliers.
 
With the above short summary at hand, we now give the relevant
notation and equations that will be used later.  For further details and
derivations of the equations below, we refer the reader to
\cite{hough04}.  
%Let $N$ be the number of SFTs, $\Tcoh$ the
%time-baseline of each SFT and $M$ the number of uniformly spaced data
%points in the time domain from which the SFT is constructed.  If the
%time series is denoted by $x_j$ ($j=0\ldots M$), then our convention
%for the discrete Fourier transform of $\{x_j\}$ is 
%
%\begin{equation} 
%\tilde{x}_k = \Delta t \sum_{j=0}^{M-1}x_j e^{-2\pi ijk/M}
%\end{equation}
%
%where $k=0,1\ldots (M-1)$, and $\Delta t = \Tcoh/M$.  
%For $0\leq k \leq \lfloor M/2 \rfloor$, the frequency index $k$ corresponds 
%to a
%physical frequency of $f_k= k/\Tcoh$ with $\lfloor .\rfloor$ denoting
%the integer part of a given real number.  The values 
%$\lfloor M/2 \rfloor < k \leq M-1$ correspond to negative frequencies given
%by $f_k = (k-M)/\Tcoh$.  We require that $\Tcoh$ is small enough so
%that the signal does not shift by more than, say, half a frequency bin
%within this time duration.  For frequencies of $\sim 300\,$Hz, this
%restricts $\Tcoh$ to be smaller than $\sim 60$min \cite{hough04}.  In
%this paper, we work with SFTs for which $\Tcoh = 1800$s.  

Frequency bins are selected by setting a threshold $\rho_\th$ on the
normalized power $\rho_k$ defined as
\begin{equation} \label{eq:normpower}
\rho_k = \frac{2|\tilde{x}_k|^2}{\Tcoh S_n(f_k)} \,,
\end{equation}
where ${\tilde{x}_k}$ is the discrete Fourier transform of the data,
the frequency index $k$ corresponds to a physical frequency of $f_k= k/\Tcoh$,
and $S_n(f_k)$ is the single sided power spectral density of the
detector noise.
%; see Appendix \ref{sec:bias} for our method of estimating 
%$S_n(f)$.  
The $k^{th}$ frequency bin is selected if $\rho_k 
\geq \rho_\th$, and rejected otherwise.  In this way, each SFT is replaced
by a collection of zeros and ones called a peak-gram.  

%Let $n$ be the final number count at a parameter space point $\xi$ obtained
%after summing the relevant bin of each peak-gram.  
Let $n$ be the  number count at a point in parameter space,  obtained
after performing the Hough transform on our data. 
Let $p(n)$ be the
probability distribution of $n$ in the absence of a signal, and
$p(n|h)$ the distribution in the presence of a signal $h(t)$.  It is
clear that $0\leq n \leq N$, and it can be shown that for stationary
Gaussian noise, $p(n)$ is a binomial distribution with mean
$Nq$ where $q$ is the probability that any frequency bin is selected:
\begin{equation}
\label{eq:binomialnosig}
p(n) = \left( \begin{array}{c} N \\ n \end{array}  \right)
q^n(1-q)^{N-n}\,.
\end{equation}
For Gaussian noise in the absence of a signal, it is easy to show that
$\rho_k$ follows an exponential distribution so that $q =
e^{-\rho_\th}$.  In the presence of a signal, the distribution is
ideally also a binomial but with a slightly larger mean $N\eta$ where,
for weak signals, $\eta$ is given by  
\begin{equation}
\eta = q\left\{1+\frac{\rho_\th}{2}\lambda +
\mathcal{O}(\lambda^2)  \right\}\,.
\end{equation}
$\lambda$ is the signal to noise ratio within a single SFT, and
for the case when there is no mismatch between the signal and the
template: 
\begin{equation} \label{eq:lambda}
\lambda = \frac{4|\tilde{h}(f_k)|^2}{\Tcoh S_n(f_k)}\,,
\end{equation}
with $\tilde{h}(f)$ being the Fourier transform of the signal $h(t)$.
The approximation that the distribution in the
presence of a signal is binomial breaks down for reasonably strong
signals.  This is due to possible non-stationarities in the
noise, and the amplitude modulation of the signal which
causes $\eta$ to vary from one SFT to another. 

Candidates in parameter space are selected by setting a threshold
$n_\th$ on the number count.  The false alarm and
false dismissal rates for this threshold are defined respectively in
the usual way: 
\begin{equation}
\alpha = \sum_{n=n_\th}^{N} p(n) \,,\qquad
\beta = \sum_{n=0}^{n_\th-1}p(n| h)\,. 
\end{equation}
We choose the thresholds $(n_\th,\rho_\th)$ based on the
Neyman-Pearson criterion of minimizing $\beta$ for a given value of
$\alpha$.  It can be shown \cite{hough04} that this criteria leads,
in the case of weak signals, large $N$, and Gaussian stationary noise, to
$\rho_\th \approx 1.6$.  This corresponds to $q \approx
0.20$, {\it i.e.}, we select about $20\%$ of the frequency bins from each
SFT. This value of $\rho_\th$ turns out to be independent of the
choice of $\alpha$ and signal strength. Furthermore, $n_\th$ is also
independent of the signal strength and is given by: 
\begin{equation}
n_\th = Nq +
\sqrt{2N q(1-q)}\,\textrm{erfc}^{-1}(2\alpha)\,,
\label{eq:nth} 
\end{equation}
where $\textrm{erfc}^{-1}$ is the inverse of the complementary error 
function.  These values of the thresholds lead to a false dismissal 
rate $\beta$ which is given in \cite{hough04}.  
The value of $\beta$ of course depends on the signal strength, and on the 
average, the weakest signal which will cross the above
thresholds at a false alarm rate $\alpha$ and false dismissal
$\beta$ is given by
\begin{equation} \label{eq:sensitivity}
h_0 = 5.34 \frac{\mathcal{S}^{1/2}}{N^{1/4}}\sqrt\frac{S_n}{\Tcoh}\,,
\end{equation}
where 
\begin{equation}
\mathcal{S} = \textrm{erfc}^{-1}(2\alpha) + \textrm{erfc}^{-1}(2\beta)\,.
\label{eq:S}
\end{equation}
Equation (\ref{eq:sensitivity}) gives the smallest signal which can be
detected by the search, and is therefore a measure of the sensitivity
of the search.  
%The sensitivity can be improved either by increasing the coherent 
%time baseline $\Tcoh$ or by incresing  the amount or data to analyze.

%The quantity $\mathcal{S}$ can be taken as a measure of the
%statistical sensitivity of the Hough transform.  This is because
%$\mathcal{S}$ is closely related to the more common definition of
%statistical sensitivity: $s = 1 - \alpha - \beta$.  In particular,
%using the properties of the error function, it can be shown that
%$\mathcal{S} = 0$ when $s = 0$.  At fixed detector noise $S_n$, the larger we 
%want $\mathcal{S}$ to be, the
%larger the value of $h_0$ and the lower the sensitivity
%of the Hough search.  

%%%%%%%%%%%%%%%%%%%%%%%%%%%%%%%%%%%%%%%%%%%%%%%%%%%%%%%%%%%%%%%%%%%%%%%%%%%%%%%%%%%%%%%%%%%%%%
\section{The search}
\label{sec:searchresults}

%Blah

%%%%%%%%%%%%%%%%%%%%%%%%%%%%%%%%%%%%%%%%%%%%%%%%%%%%%%%%%%%%%%%%%%%%%%%
\subsection{The SFT data}
\label{subsec:sftdata}

%The method used to analyze these data is the one
%described in \cite{hough04} as the {\it Hough search with
%non-demodulated data}. 
The input data to our search is a collection of calibrated SFTs 
with a time baseline $\Tcoh$ of 30 minutes.
While a larger value of $\Tcoh$ leads to better sensitivity, this
time baseline can not be made arbitrarily large 
because of the frequency drift caused by the Doppler effect (and also
the spin-down); we would like the signal power of a putative signal to be
concentrated in less than half the frequency resolution
$1/\Tcoh$.  It is shown in \cite{hough04} that at $300\,$Hz, we could
ideally choose $\Tcoh$ up to $\sim60\,$min.  On the other hand, we
should be able to find a significant number of such data stretches
during which the interferometers are in lock, the noise
is stationary, and the data are labeled satisfactory according to 
certain data quality requirements.   Given the duty cycles of the
interferometers during S2 
and the non-stationarity of the noise floor, it turns out that
$\Tcoh=30\,$min is a good compromise which satisfies these
constraints.  By demanding the data in each 30 minutes stretch to be 
continuous (although there could be gaps in 
between the SFTs) the number $N$ of 
SFTs available for L1 data is 687, 1761 for H1 and 1384 for H2, reducing the
nominal duty cycle for this search. 

The SFT data are calibrated in the frequency domain by constructing a
response function $R(f,t)$ that acts on the error signal of
the feedback loop used to control the differential length of the interferometer  
arms.  %The differential arm length strain is then given by %$= R(f,t)\,q(f)$. 
The response function $R(f,t)$ varies in time,  primarily due
to changes in the amount of light stored in the Fabry-Perot cavities
of the interferometers.   
During S2, changes in the response were computed every 60 seconds,
and an averaging procedure was used
to estimate the response function used on each SFT.
%Interferometer calibration for the S2 run is described in detail in
%\cite{S2caldoc}.
%The re-construction of the strain from the output of the
%interferometer feedback loop is referred to as the calibration. 
% 
% \be
% h(f)=R(f,t)\,q(f).
% \label{hf}
% \ee
%
The SFTs are windowed and high-pass filtered as described in 
Sec.~IV~C~1 of \cite{S1-CW}. No
further data conditioning is applied, although the data are known to
contain many spectral disturbances, including the $60\,$Hz power
line harmonics and the thermally excited violin modes of test mass 
suspension wires.

%%%%%%%%%%%%%%%%%%%%%%%%%%%%%%%%%%%%%%%%%%%%%%%%%%%%%%%%%%%%%%%%%%%%%%%%%%%%%%%%%%%%%%%%%%%
\subsection{The parameter space}
\label{subsec:parameters}

This section describes the portion of parameter space
$(f_0,\dot{f},\mathbf{n})$ we search over, and the resolution of our grid. 
Our template grid is not based on a metric calculation (as in
e.g. \cite{BCCS, BC00}), but rather on a cubic grid which covers the
parameter space as described below. Particular features of this grid
are used to increase computational efficiency as described in
Sec.~\ref{subsec:pipeline}.    

We analyze the full data set from the S2 run with a total observation 
time $\Tobs \sim 5.1\times 10^6$ sec. The exact value of $\Tobs$ is 
different for the three LIGO interferometers \footnote{$\Tobs$ is the time
from the first data to the last data of the SFTs used from the S2 run.
Because the data are not continuous $\Tobs > N\Tcoh$ and it is also different
for the three detectors.}.  
We search for isolated neutron star signals 
in the frequency range $200$--$400\,$Hz with a frequency resolution
\be
\delta f = \frac{1}{\Tcoh} = 5.556\times 10^{-4}\,\textrm{Hz}\,.
\ee
The choice of the range $200$--$400\,$Hz for the analysis is motivated 
by the low noise level, and therefore our ability to set
the best upper limits for $h_0$, as seen from Fig.
\ref{fig:S2sensitivity}.
  
The resolution $\delta\dot{f}$ in the space of first spin-down
parameters is given by the smallest value  of $\dot{f}$ for which the
intrinsic signal frequency does not drift by more than a single
frequency bin during the total observation 
time \footnote{The spin-down resolution depends on the total observation time,
and this turns out to be $-1.10508\times 10^{-10}\,\textrm{Hz-s}^{-1}$ for L1,
 $-1.09159\times 10^{-10}\,\textrm{Hz-s}^{-1}$ for H1, and
 $-1.10388\times 10^{-10}\,\textrm{Hz-s}^{-1}$ for H2.}: 
\be \label{eq:deltafk}
\delta \dot f = \frac{\delta f}{\Tobs}=\frac{1}{\Tobs\Tcoh}
\sim 1.1\times 10^{-10}\, \textrm{Hz-s}^{-1}\,.
\ee
We choose the range of values $-\dot f_\max \le \dot f\le 0$, where the
largest spin-down parameter $\dot f_\max$ is about 
$1.1\times 10^{-9}\,\textrm{Hz-s}^{-1}$. This yields eleven spin-down values  
for each intrinsic frequency. In other words, we look for neutron stars
whose spin-down age is at least $\tau_\min = {\hat f}/{\dot f_\max}$.
This corresponds to a minimum spin-down age of $5.75\times 10^{3}\,$yr at
$200\,$Hz, and $1.15\times 10^{4}\,$yr at $400\,$Hz.  These values of
$\dot{f}_\max$ and $\tau_\min$ are such that all known pulsars have a
smaller spin-down rate than $\dot{f}_\max$ and, except for a few
supernova remnants, all of them have a spin-down age significantly
greater than the numbers quoted above.  With these values of
$\tau_\min$, it is easy to see that the second spin-down parameter can
be safely neglected; it would take about $10\,$yr for the largest
second spindown parameter to cause a frequency drift of half a
frequency bin.  

As described in \cite{hough04},  
for every given time, value of the intrinsic frequency $f_0$ 
and spin-down $\dot{f}$, the set of
sky-locations $\mathbf{n}$ consistent with a selected frequency ${f}(t)$ 
corresponds
to a constant value of $\mathbf{v}\cdot\mathbf{n}$ given by
(\ref{eq:master}).  This is a circle in the celestial sphere.  It can
be shown %\cite{hough04} 
that every  frequency bin of width
$\delta f$ corresponds to an annulus on the celestial sphere whose
width is at least 
\begin{equation} \label{eq:annuluswidth}
(\delta \theta)_\min= \frac{c}{v}\frac{\delta f}{\hat{f}} \, ,
\end{equation}
with $v$ being the magnitude of the average velocity of the detector
in the SSB frame. %during the 30 minutes interval of the corresponding SFT. 

The resolution $\delta\theta$ in sky positions is chosen to be
 frequency dependent,
%with the number of templates increasing with frequency, 
being at most
$\delta\theta= \frac{1}{2}(\delta \theta)_\min$.
To choose the template spacing only, we use a constant value of 
$v/c$ equal to $1.06\times 10^{-4}$.
% where $(\delta\theta)_\min$ is
%given in Eq.~(\ref{eq:annuluswidth}).  
This yields: 
\begin{equation}
  \delta\theta  =  9.3\times
  10^{-3}\,\textrm{rad}\times
\left(\frac{300\,\textrm{Hz}}{\hat{f}}\right) \, .
\label{eq:minannulus} 
\end{equation}
This resolution corresponds to approximately $ 1.5\times 10^{5}$ sky
locations for the whole sky at $300\,$Hz. 
For that, we break up the sky
into 23 sky-patches of roughly equal area and, by means of the
stereographic projection, we map each portion to a
plane, and  set a uniform  grid with
spacing $\delta \theta$ in this stereographic plane.
% and are finally projected back on to the celestial sphere.  
The stereographic projection maps circles in the celestial
sphere to circles in the plane thereby mapping the annuli in the
celestial sphere, described earlier, to  annuli in the
stereographic plane.  We ensure that the dimensions of each sky-patch
are sufficiently small so that the distortions produced by the
stereographic projection are not significant.
This is important to ensure that the number of points  needed to cover the
the full sky is not much larger than if we were using exactly the 
frequency resolution given by Eq.~(\ref{eq:minannulus}).
%that the coverage on the sky is 
% nearly constant density. {\bf check last statement}.

This adds up to a total number of templates per $1\,$Hz band at $200\,$Hz
$\sim 1.9 \times 10^{9}$ while it increases up to $7.5 \times 10^{9}$
at $400\,$Hz.

%%%%%%%%%%%%%%%%%%%%%%%%%%%%%%%%%%%%%%%%%%%%%%%%%%%%%%%%%%%%%%%%%%%%%%%%%%%%%%%%%%%%%%
\subsection{The implementation of the Hough transform}
\label{subsec:pipeline}

This section describes in more detail the implementation 
of the search pipeline which was
summarized in Sec.~\ref{subsec:houghbasics}.   
%\textcolor{red}{\sc To be shorten}
%As described in \cite{hough04},
The first step in this semi-coherent Hough search is to select
frequency bins from the SFTs and construct the peak-grams.
As mentioned in Sec.~\ref{subsec:houghbasics}, our criteria for
selecting frequency bins is to set a threshold of $1.6$ on the
normalized power (\ref{eq:normpower}), thereby selecting about $20\%$
of the frequency bins in every SFT.  

The power spectral density  $S_n$ appearing in
Eq.~(\ref{eq:normpower}) is estimated by means of a running median 
applied to the periodogram of each individual SFT.
The window size we employ for the running median
is $w=101$ corresponding to $0.056\,$Hz 
\footnote{The window size $w$ should be as small as possible in 
order to track the noise floor accurately.  On the other hand, the 
statistical errors in the value of the running median are smaller 
when a large window size is chosen. We have found that $w=101$ is a 
good compromise.}. 
The running median
is a robust method to estimate the noise floor \cite{mohanty02b,
  mohanty02a,badri} which has the virtue of discarding outliers which appear
in a small number of bins, thereby providing
an accurate estimate of the noise floor in the presence of
spectral disturbances and possible signals. The use of
the median (instead of the mean) to estimate the power spectral density
 introduces a minor technical complication (see appendix \ref{sec:bias}
 for further details).

The next step is to choose a tiling of the sky. As described before, 
we break up the sky
into 23 patches, of roughly equal area. %, which overlap.
By means of the stereographic projection, we map each portion 
to a two dimension plane and set a uniform grid  with a
resolution $\delta \theta$ in this plane.
All of our calculations are performed on this stereographic plane, and
are finally projected back on to the celestial sphere.  

In our implementation of the Hough transform, we treat
sky-positions separately from frequencies and spin-downs. In
particular, we do not obtain the Hough histogram over the entire
parameter space in one go, % directly, 
but rather for a given sky-patch, a search frequency $f_0$ and 
a  spin-down $\dot{f}$ value.
These are the so-called \emph{Hough maps} (HMs). Repeating this for
every set of frequency and spin-down parameters and the different sky-patches
we wish to search over, we obtain a number of HMs. The collection of
all these HMs represent our final histogram in parameter space.

\begin{figure}
  \begin{center}
  \includegraphics[height=5cm,width=8.5cm]{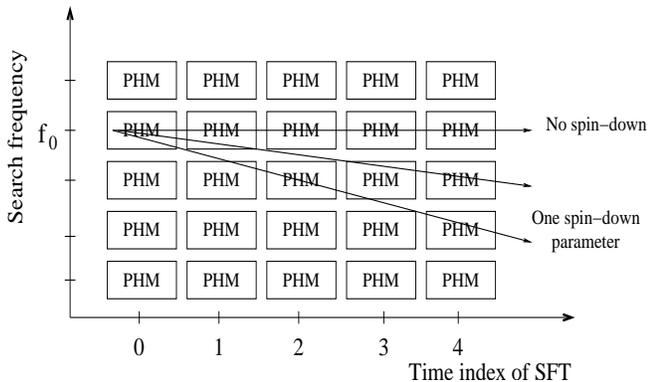}
  \caption{A partial Hough map (PHM) is a histogram in the
  space of sky locations obtained by performing the 
  Hough transform on a single SFT
  %$(\alpha,\delta)$ plane constructed from all the frequencies
  %selected at a \emph{given} time 
  and for a given value
  of the instantaneous frequency.  A total Hough map is
  obtained by summing over the appropriate PHMs.  The PHMs to be
  summed over are determined by the choice of spin-down parameters
  which give a trajectory in the time-frequency plane.  
  %For example, a single spin down parameter will give a straight line as
  %shown is the figure while two spin-down parameters will lead to a
  %parabola.
  }\label{fig:sumphm}
  \end{center}
\end{figure}
\begin{figure}
  \begin{center}
  \includegraphics[width=8cm]{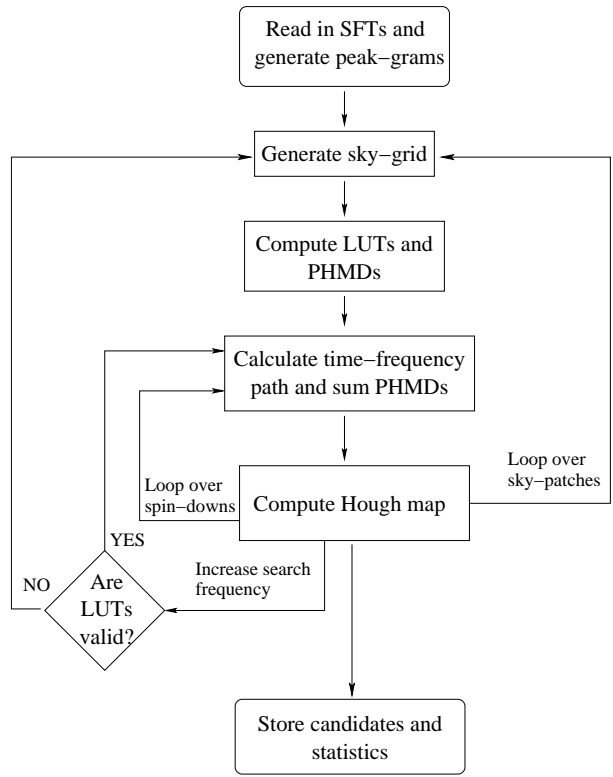}
  \caption{The schematic of the analysis pipeline. The input data are the 
  SFTs and the search parameters.
  The first step is to select frequency bins from the SFTs and generate the 
  peak-grams. Then, the Hough transform is computed  for the different sky
  patches, frequencies and spin-down values, thus producing the different
  Hough maps. The search uses LUTs that are computed for a given 
  tiling of the sky-patch. % and the corresponding times of the SFTs. 
  The sky grid is frequency dependent, but it is fixed for the frequency 
  range in which the LUTs are valid. Then, a collection of PHMDs is built, and 
  for each
  search frequency $f_0$ and given spin-down $\dot f$, the trajectory in
  the time-frequency plane is computed and the Hough map obtained 
  by summing and
  integrating the corresponding PHMDs. The code  loops over frequency and
  spin-down parameters, updating the sky grid and LUTs whenever required.
  Statistical analyses are performed on each map in order to reduce the
  output size. 
  }
  \label{fig:S2houghPipeline} 
  \end{center}
\end{figure}

The HMs could be produced by using a ``brute force'' approach, {\it i.e.},
using all the peaks in the time-frequency plane. But there is an alternative
way of constructing them. Let us define a \emph{partial Hough map} (PHM)
as being a Hough histogram, in the space of sky locations,
obtained by performing the
Hough transform using the peaks from a \emph{single} SFT and for a 
\emph{single} value of the
intrinsic signal frequency and no spin-down.
This PHM therefore consists of only zeros and ones, {\it i.e.},
the collection of the annuli corresponding to all peaks present in a 
single peak-gram.
Then each HM can be obtained  by summing the appropriate PHMs 
produced from different SFTs. %, as shown in Fig.~\ref{fig:sumphm}.  
If we add PHMs constructed by using the same intrinsic frequency, then
the resulting HM refers to the same intrinsic frequency and no spin-down.
But note that the effect of a spin-down in the signal is the same as having
a time varying intrinsic frequency. This suggests a strategy to re-use
PHMs computed for different frequencies at different times in order to
compute the HM for a non-zero spin-down case.

%Each HM is obtained  
%%by summing the appropriate \emph{partial Hough maps} (PHMs).  A PHM is
%a Hough histogram in the space of sky locations, obtained by performing the
%Hough transform on a \emph{single} SFT and for a \emph{single} value of the
%intrinsic signal frequency $\hat{f}(t)$ at the time corresponding
%to the mid point of the SFT; %(Eq.~\ref{eq:fhat}); 
%it therefore consists of only
%zeros and ones.  

Given the set of PHMs, the  HM for a
given search frequency $f_0$ and a given spin-down $\dot{f}$ is obtained as
follows:  using Eq.~(\ref{eq:fhat}) calculate the trajectory
$\hat{f}(t)$ in the time-frequency plane corresponding to $f_0$ and
$\dot{f}$.  If the mid time stamps of the SFTs are $\{t_i\}$ ($i=1\ldots
N$), calculate $\hat{f}(t_i)$ and find the frequency bin that it lies
in; select the PHM corresponding to this frequency bin.  Finally, add
all the selected PHMs to obtain the Hough map.  This procedure is shown in
Fig.~\ref{fig:sumphm}.  

This approach saves computations because it recognizes that the same sky
locations can refer to different values of frequency and spin-down, and
%$f_0$ and $\dot{f}$, and
avoids having to re-determine such sky locations more than once.
Another  advantage of proceeding in this fashion
is that %PHMs can be reused for different search frequencies and spin-down values and
we can use  \emph{look up tables} (LUTs) to construct the PHMs.  
The basic problem to construct the PHMs 
is that of drawing the annuli on the celestial sphere,
or on the corresponding projected plane. 
The algorithm we use based
on LUTs has proved to be more efficient than other methods we have studied,
and this strategy is also employed by other groups \cite{houghvirgo}.

A LUT is an array containing the list of borders of all the
possible  annuli, for a given value of $\mathbf{v}$ and $\hat{f}$,
 clipped on the sky-patch we use. 
Therefore it contains the coordinates of the points
belonging to the borders  that intersect the sky-patch,
 in accordance to the tiling we use,
together with
information to take care of edge effects.
As described in \cite{hough04}, it turns out that the annuli are
relatively insensitive to the value of the search frequency and, once
a LUT has been constructed for a particular 
frequency, it can be re-used for a large number of
neighboring  frequencies thus allowing for computational savings.
 The value of $\mathbf{v}$  used to construct the LUTs corresponds
 to the average velocity of the detector in the SSB frame during the 30 minutes
 interval of the corresponding SFT.
%Note that an approximation takes place when selecting the PHM to be added
%to construct a Hough map. The PHMs are computed only for 
%frequency values at integer multiples of  $1/\Tcoh$. Given a trajectory
% in the time-frequency plane corresponding to $f_0$ and
%$\dot{f}$, then $\hat f(t)$ are rounded to the nearest resolved frequency.

In fact, the code is further sped up by using 
\emph{partial Hough map derivatives} (PHMDs) instead of the PHMs, 
in which only the borders of the annuli are indicated.
A PHMD  consists of only ones, zeros, and minus ones, 
in such a way that by integrating appropriately over the different sky 
locations one recovers the corresponding PHM. This integration is 
performed at a later stage, and just once, after summing the appropriate 
PHMDs, to obtain the final Hough map.

In the pipeline, we loop over frequency and spin-down 
values, taking care to update the set of PHMDs currently used,
 and  checking the validity of the LUTs.
As soon as the LUTs are no longer valid, the code
recomputes them again together with
 the sky grid.
Statistical analyses are performed on the Hough maps in order to compress
the output size. These include finding the maximum, minimum, mean and
standard deviation of the number counts for each individual map, the
parameters of the loudest event, and also of all the candidates above 
a certain threshold. We also
record the maximum number count per frequency analyzed, maximized
over all spin-down values and sky locations, and a
histogram of the number counts for each $1\,$Hz band.
The schematic search pipeline is shown in Fig. \ref{fig:S2houghPipeline}.

As a technical implementation detail, the search is performed by 
dividing  the $200\,$Hz frequency band into smaller bands of $1\,$Hz 
and  distributed using Condor \cite{condor} on 
a computer cluster.
%the Merlin cluster at AEI \cite{merlin}. 
Each CPU analyzes a 
different $1\,$Hz band using the same pipeline (as described
in Fig.~\ref{fig:S2houghPipeline}). The code itself takes care to read
in the proper frequency band from the SFTs.  
This includes the search band plus an extra interval to accommodate for 
the maximum Doppler shift, spin-down, and
 the block sized used by the running median.
 %, in order to avoid edge effects.
 The analysis described here was carried out on the 
 Merlin cluster at AEI \cite{merlin}.
The full-sky search for the entire S2 data from the 
three detectors distributed 
on 200 CPUs on Merlin lasted less than half a day.

The software used in the analysis is available in the LIGO Scientific
Collaboration's CVS archives at 
{\tt http://www.lsc-group.phys.uwm.edu/cgi-bin/cvs/ viewcvs.cgi/?cvsroot=lscsoft}, 
together with a suite of test programs, especially for 
visualizing the Hough LUTs. The full search pipeline
has also been validated by comparing the results with 
independently written code that implements a less efficient 
but conceptually simpler  approach, {\it i.e.}, for each point in parameter 
space  $(f_0,\dot{f},\mathbf{n})$, it finds the corresponding pattern 
in the time-frequency plane
and sums the corresponding selected frequency bins.

%%%%%%%%%%%%%%%%%%%%%%%%%%%%%%%%%%%%%%%%%%%%%%%%%%%%%%%%%%%%%%%%%%%%%%%%%%%%%%%%
\subsection{Number counts from L1, H1 and H2}
\label{subsec:numberCounts}

%We use the Hough transform to perform a full sky search for isolated neutron
%stars between 200 Hz and 400
%Hz with one spin down parameter, using the full data set from the S2 run.
% 
%The resolution in parameter space corresponds to the one 
%described in subsection \ref{subsec:parameters}.
%The total number of templates analyzed in a one
%Hz band at 200 Hz is roughly $1.9 \times 10^{9}$ while it increases up
%to $7.5 \times 10^{9}$ at 400 Hz.

\begin{figure}
  \begin{center}
  \includegraphics[height=8cm, width=8cm]{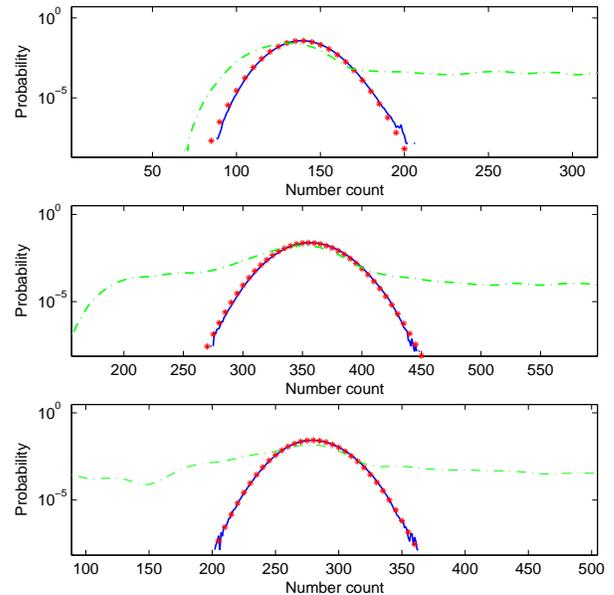}
  \caption{Top: Graph of the L1 number-count discrete 
  probability distribution: 
  the solid curve corresponds to the number count distribution obtained
  for the 
  band between $206$--$207\,$Hz, the dash-dot curve  to the number count
  distribution obtained for the $343$--$344\,$Hz band, that contains violin
  modes, and in  asterisks  the theoretical expected binomial distribution
  for 687 SFTs and a peak selection probability of  $20\%$.
  Middle: the H1 number-count distribution for 1761 SFTs.
  Bottom: the H2 number-count distribution for 1384 SFTs.}
  \label{fig:histo}
  \end{center}
\end{figure}

%
%\begin{figure}
%  \begin{center}
%  \includegraphics[height=5.3cm]{H1histo_206_343}
%  \caption{Graph of the H1 number-count distribution, as in figure 
%  \ref{fig:histoL1}
%  but for 1761 SFTs.}
%  \label{fig:histoH1}
%  \end{center}
%\end{figure}
%

%\begin{figure}
%  \begin{center}
%  \includegraphics[height=5.3cm]{H2histo_206_343}
%  \caption{Graph of the H2 number-count distribution, as in figure 
%  \ref{fig:histoL1}
%  but for 1384 SFTs.}
%  \label{fig:histoH2}
%  \end{center}
%\end{figure}
%

In the absence of a signal, the distribution of the Hough number count 
ideally is a binomial distribution.
Environmental and instrumental noise sources can excite
the optically sensed cavity length, or get into the output signal in some
other way, and  show up as  spectral disturbances,
such as lines. % -- of course not all spectral disturbances are of this nature. 
If no data conditioning is applied, line interference
can produce an excess of number counts in the 
Hough maps and mask signals from a wide area in the sky.
Fig. \ref{fig:histo} shows the comparison of
the theoretical binomial distribution
Eq.~(\ref{eq:binomialnosig}) with the distributions that we 
obtain experimentally in two bands: $206$--$207\,$Hz and
$343$--$344\,$Hz.  
The first 
band contains very little spectral disturbances while the second band
contains some violin modes.
As shown in  Fig.~\ref{fig:histo},
the Hough number count follows the expected binomial 
distribution for the {\it clean} band while it diverges from the expected 
distribution in the presence of strong spectral disturbances, such
as the violin modes in this case. We have verified good agreement
in several different frequency bands
that were free of strong spectral disturbances. 
 
The sources of the  disturbances present in the S2 data 
are mostly understood. They consist 
of calibration lines, broad $60\,$Hz power line harmonics,  
multiples of $16\,$Hz due to the data acquisition system, and a number of
mechanical resonances, as for example  
the  violin modes of the mirror 
suspensions \cite{S2caldoc,S2-lines}. 
%{\bf make sure 
 % that all the refs to web pages are removed and the relevant docs 
 % submitted to the DCC and these as cited}.
The $60\,$Hz power lines are rather broad, with a width of about $\pm 0.5\,$Hz,
while the calibration lines and the $16\,$Hz data acquisition lines are confined
to a single frequency bin. 
A frequency comb is also present in the data, having fundamental frequency 
at $36.867\,$Hz for L1,
$36.944\,$Hz for H1 and $36.975\,$Hz for H2,  some of them  accompanied 
with side lobes at 
about $0.7\,$Hz, created by up-conversion of the 
pendulum modes of some core optics, but these were only present 
(or at least prominent) in H1 and H2. The sources of these combs were 
synthesized
oscillators used for phase modulation that were later replaced by crystal 
oscillators.
In addition to the above disturbances,  we also observe a large number of
multiples of $0.25\,$Hz. While this comb of lines is strongly suspected 
to be instrumental, its physical origin is unknown.
In Table~\ref{tab:S2knownlines} we summarize the list of known spectral 
disturbances in the three interferometers during the S2 run.

%
%\begin{table}
%\begin{tabular}{c|ll||c|ll}\hline
%           &                &                 & & &\\ 
%  Detector &   $f$ (Hz)     & $\Delta f$ (Hz) & Detector  & $f$ (Hz)
%  & $\Delta f$ (Hz)\\
%           &                &                 & & &\\
%\hline \hline
% & & & & & \\ 
%  All      &  $60^\star$    & $\pm 0.5$       &  H1  &  $345$   & $\pm 3.0$       \\
%           &  $0.25^\star$  & 0               &      &  $36.944^\star$  & $0$             \\
%           &  $16^\star$    & 0               & & & \\ 
% & & & & & \\
%\hline
% & & & & & \\
%  H2       &  $346.5$       & $\pm 4.5$       & L1   &  $345$       & $\pm 3.0$    \\
%           &  $36.975^\star$& $0$             &      &  $36.867^\star$
%  & $0$ \\
% & & & & & \\
%\hline
%\end{tabular}\\
%$\star$ indicates all higher harmonics are present
%\caption{List of known spectral disturbances in the three
%  interferometers during the S2 run \textbf{Include more lines??
%   All of them should be here!!!!!!!!!!!!!!!!!!!!!!!!!!!!!
%   with the exact width we used!!!!!!!!!!!!!!}.  } 
%\label{tab:S2knownlines}
%\end{table}

%%%%%%%%%%%%%%%%%%%%%%%%%%%%%%%%%%%%%%%%5

\begin{table}
\begin{tabular}{cccccc}\hline
\multicolumn{2}{c}{L1}& \multicolumn{2}{c}{H1}& \multicolumn{2}{c}{H2}\\
 $f$ (Hz) & $\Delta f$ (Hz) & $f$ (Hz) & $\Delta f$ (Hz)& $f$ (Hz) & $\Delta f$ (Hz)\\
 \hline \hline  
  $0.250^\star$  & 0.00 &   $0.250^\star$  & 0.00 &   $0.250^\star$  & 0.00 \\
 $16.000^\star$  & 0.00 &  $16.000^\star$  & 0.00 &  $16.000^\star$  & 0.00 \\ 
 $60.000^\star$  & 1.00 &  $60.000^\star$  & 1.00 &  $60.000^\star$  & 1.00 \\
         221.200 & 0.01 &  221.665 & 0.01 & 221.850 & 0.02 \\
         258.080 & 0.04 &  257.875 & 0.02 & 258.830 & 0.01 \\
	 294.935 & 0.01 &  258.610 & 0.02 & 295.070 & 0.02 \\	 
	 331.810 & 0.01 &  259.340 & 0.02 & 295.670 & 0.02 \\ 
         345.000 & 6.00 &  294.820 & 0.01 & 295.800 & 0.04 \\ 
	 368.670 & 0.02 &  295.560 & 0.00 & 295.930 & 0.02 \\ 
	         &      &  296.300 & 0.00 & 296.530 & 0.02 \\
	         &      &  331.790 & 0.00 & 323.300 & 0.00 \\
		 &      &  332.490 & 0.00 & 323.870 & 0.04 \\
		 &      &  333.200 & 0.00 & 324.000 & 0.04 \\
		 &      &  335.780 & 0.14 & 324.130 & 0.04 \\
		 &      &  336.062 & 0.00 & 324.700 & 0.00 \\
	         &      &  339.000 & 0.02 & 332.800 & 0.00 \\
	         &      &  339.720 & 0.02 & 335.120 & 0.02 \\
	         &      &  345.000 & 6.00 & 335.590 & 0.02 \\
	         &      &  365.500 & 0.02 & 341.615 & 0.01 \\
	         &      &  368.690 & 0.00 & 346.500 & 9.00 \\
	         &      &  369.430 & 0.00 & 349.202 & 0.00 \\  
	         &      &  370.170 & 0.01 &   &   \\ 
\hline
\end{tabular}\\
$\star$ indicates all higher harmonics are present
\caption{List of known spectral disturbances in the three
  interferometers during the S2 run  used as a frequency veto
  in the 200--$400\,$Hz band.  $f$ refers to the central frequency
  and  $\Delta f$ to the full width of the lines. 
  Lines denoted with $\Delta f=0.0\,$Hz
   are those in which the line width is much smaller than the associated
   maximum Doppler broadening of the line. This ranges from 
  $\sim 0.04\,$Hz at a frequency of $200\,$Hz up to 
   $\sim 0.08\,$Hz at  $400\,$Hz.    } 
\label{tab:S2knownlines}
\end{table}

\begin{figure}
  \begin{center}
  \includegraphics[height=7cm]{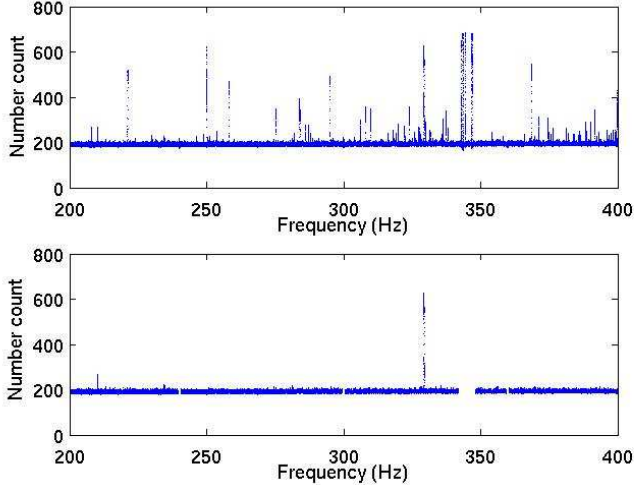}
  \caption{Graph of the L1 maximum number count $n_k^\star$ for every
  analyzed frequency $f_k$, maximized over all spin-down values and
  sky locations. The top figure
  corresponds to the raw output from the Hough transform in which many
  outliers are clearly visible. The bottom figure corresponds to the same
  data after vetoing the  frequency bands contaminated
  by known instrumental noise. See App.~\ref{app:outliers} for details on outliers.}
  \label{fig:S2L1ncMax}
  \end{center}
\end{figure}
\begin{figure}
  \begin{center}
  \includegraphics[height=7cm]{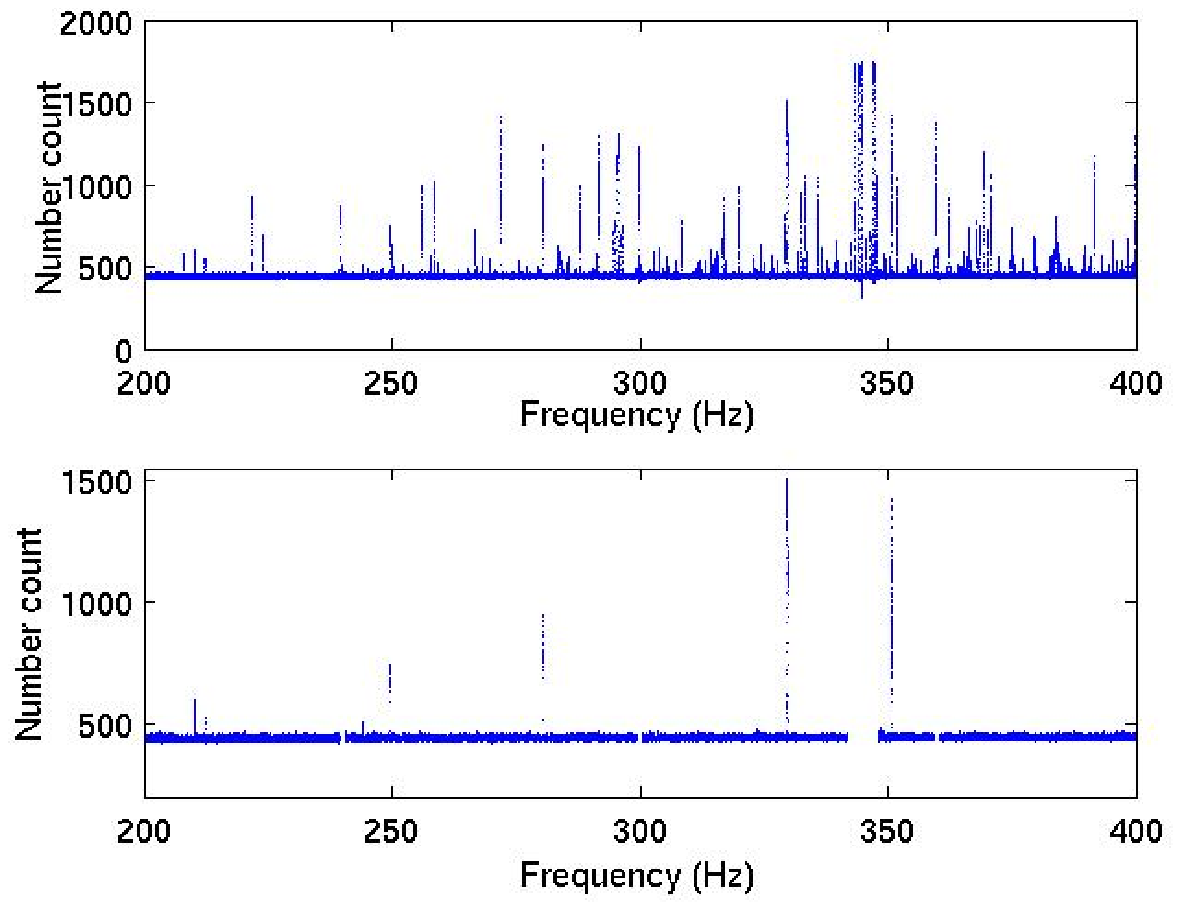}
  \caption{Graph of the H1 maximum number count $n_k^*$ versus frequency $f_k$
  as in Fig. \ref{fig:S2L1ncMax}.}
  \label{fig:S2H1ncMax}
  \end{center}
\end{figure}
\begin{figure}
  \begin{center}
  \includegraphics[height=7cm]{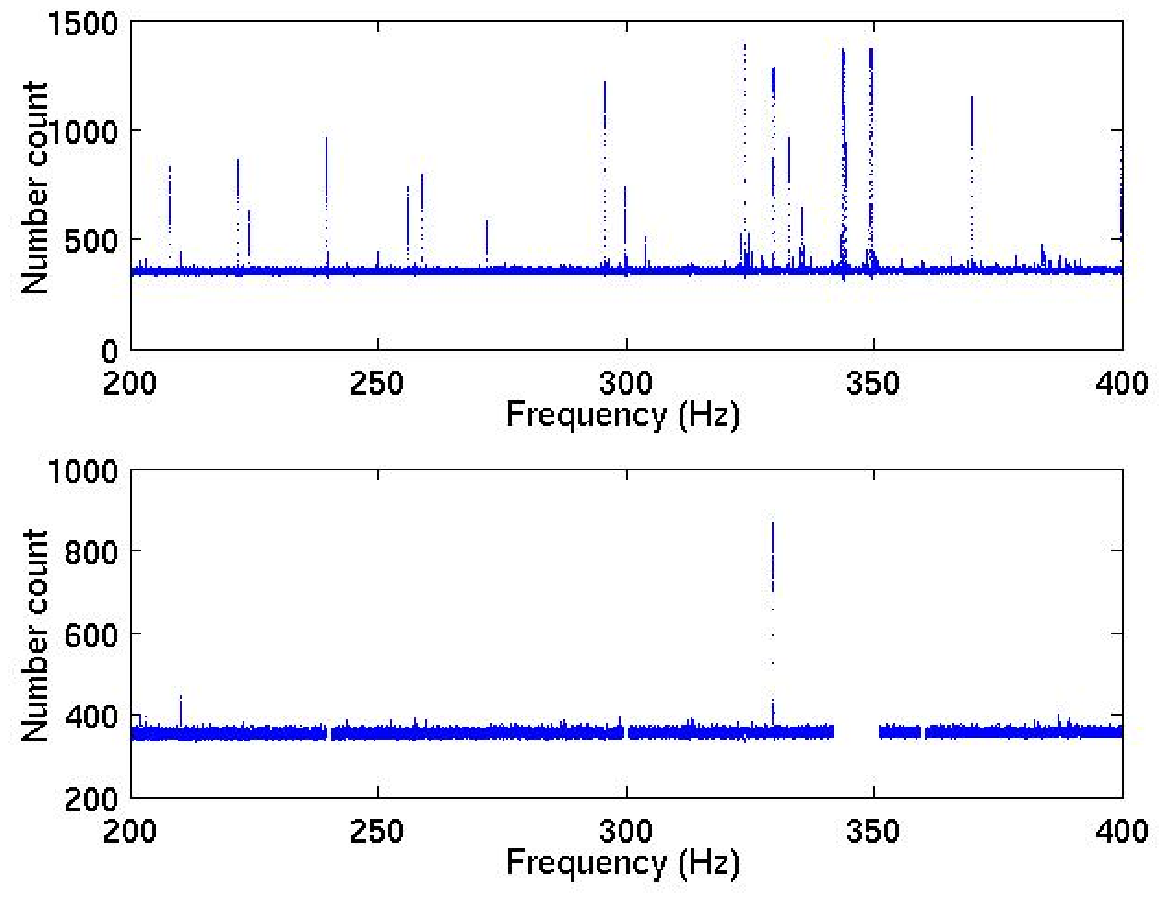}
  \caption{Graph of the H2 maximum number count $n_k^*$ versus frequency $f_k$
  as in Fig. \ref{fig:S2L1ncMax}. }
  \label{fig:S2H2ncMax}
  \end{center}
\end{figure}
\begin{figure}
  \begin{center}
  \includegraphics[height=9cm]{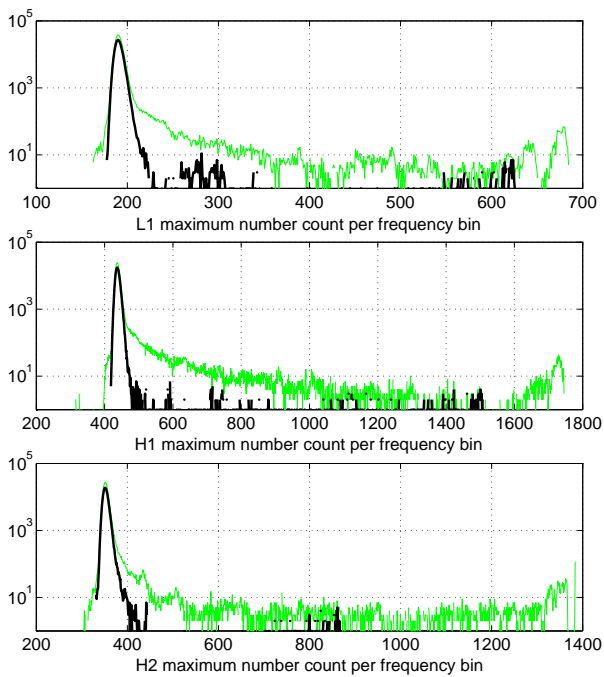}
  \caption{Histograms of the maximum  Hough number count $n_k^*$ 
  for the three detectors. The light  curve corresponds to the raw output and
  the dark thick curve corresponds to the same output after vetoing the 
  contaminated frequencies. }
  \label{fig:S2ncMax}
  \end{center}
\end{figure}

After analyzing all the data we discard the number counts 
from all those frequencies that could be contaminated by a \emph{known}
instrumental artifact. Thus, we exclude every
frequency bin which is affected by the spectral disturbances including
the maximum possible Doppler broadening of these lines; thus, for a
known spectral disturbance at a frequency $f$ and width $\Delta f$,
we exclude a frequency
range $\pm (vf/c+\Delta f/2)$ around the frequency $f$. 
We also exclude from our
analysis the frequency band 342--$348\,$Hz for L1 and H1 and 342--$351\,$Hz 
for H2 because they contain many violin modes. The net effect of this
vetoing strategy is that we consider only 67.1\% for L1,  66.8\% for H1 and
65.6\% for H2
of the full $200\,$Hz range.

Figures \ref{fig:S2L1ncMax}, \ref{fig:S2H1ncMax} and 
\ref{fig:S2H2ncMax} show the maximum Hough number count $n_k^*$ obtained
in each of the 360000 different
frequency bins  $f_k$ analyzed, maximized over all sky locations 
and spin-down values for the three detectors. In those figures we compare
the results of the search before and after applying the frequency
veto, showing how the spectral disturbances impact on the number counts.
This can be seen more clearly in Fig.~\ref{fig:S2ncMax}, where we plot
the histograms of these maximum Hough number counts $n_k^*$
before and after applying the frequency veto.

\begin{table}
\begin{tabular}{ccccccc}\hline
          &      &     & \multicolumn{2}{c}{Before Veto}
           &\multicolumn{2}{c}{After Veto} \\
           & $n_\th(\alpha_{200})$ & $n_\th(\alpha_{400})$ & $\langle n_k^*\rangle $ &
	 std($n_k^*$) & $\langle n_k^*\rangle $ & std($n_k^*$)\\ 
\hline \hline 
L1  &  188.8 &191.7 & 194.8 & 32.3 & 191.4 & 10.2 \\
H1  &  435.8 &440.3 & 452.0 & 94.0 & 439.8 & 29.5 \\
H2  &  350.6 &354.6 & 360.6 & 72.3 & 353.2 & 11.3 \\
\hline
\end{tabular}
\caption{Comparison between the statistics of the maximum Hough number count 
$n_k^*$ (before and after the frequency veto) 
and $n_\th(\alpha)$ at different false alarm rates 
$\alpha_{200}=9.5 \times 10^{-7}$ and
$\alpha_{400}=2.4 \times 10^{-7}$ for the three detectors.} 
\label{tab:S2nc}
\end{table}

These values of $n_k^\star$ obtained after removing the known outliers 
can be easily seen to be consistent with what we would expect for Gaussian 
stationary noise. Assuming that the maximum occurs only once, the expected 
value of $n_k^\star$ should be consistent with a false alarm rate of 
$1/m_{\textrm{\mbox{\tiny{bin}}}}(f_k)$, where $m_{\textrm{\mbox{\tiny{bin}}}}(f_k)$ 
is the total number of templates at a frequency $f_k$.
$m_{\textrm{\mbox{\tiny{bin}}}}(f_k)$ is frequency dependent, ranging from 
$\sim 1.1 \times 10^6$ at $200\,$Hz to $\sim 4.2 \times 10^6$ at $400\,$Hz.
%The total number of templates analyzed in this $200\,$Hz band is 
%roughly $10^{12}$.  
Thus, $n_k^\star$ should be consistent with a
false alarm rate of $\alpha_{400}=2.4 \times 10^{-7}$ at $400\,$Hz, up
to $\alpha_{200}=9.5 \times 10^{-7}$ at $200\,$Hz. 
In the case of Gaussian stationary noise, the expectation value of
$n_k^*$ should therefore be similar to $n_\th$ (defined by Eq.~(\ref{eq:nth}))
for a false alarm  $\alpha =1/m_{\textrm{\mbox{\tiny{bin}}}}(f_k)$. 
Table~\ref{tab:S2nc} compares 
the mean $\langle n_k^\star \rangle$ of the maximum Hough number count
before and after vetoing, with $n_\th(\alpha)$ at different  
false alarm rates. After vetoing, we observe
that $\langle n_k^*\rangle $ lies within the interval 
($n_\th(\alpha_{200})$, $n_\th(\alpha_{400})$), and the standard deviation
std($n_k^*$) is also greatly reduced.  This indicates the consistency
of the observed values of $n_k^\star$ with ideal noise. 

As can be seen in Fig.~\ref{fig:S2L1ncMax}, \ref{fig:S2H1ncMax} and 
\ref{fig:S2H2ncMax}, a few outliers remain after applying the frequency 
veto described above. 
%The fact that these outliers have a small false alarm makes them
%unlikely to be noise. 
The fact that these outliers have such small false alarm probabilities
makes it very unlikely they were drawn from a parent Gaussian
distribution.
However they could also be due to spectral disturbances 
(line noise) that mimic
the time-frequency evolution of a pulsar for a certain location in the sky.
If these outliers are due to gravitational signals they should
show up in the different detectors. By performing a simple
coincidence analysis in frequency, the only candidate that remains %at the
%false alarm level of $10^{-13}$ (this is equivent to a probability of $10 \%$ 
%of having one outlier in each detector),
is the one at $210.360\,$Hz. But this has been ruled out  
since it  seems to be associated to multiples of 
$70.120\,$Hz produced by a VME (VERSA module Eurocard) controller hardware 
used during S2. Since the other outliers are not coincident among
the three detectors, there is no evidence for a detection.
We refer the interested reader to appendix~\ref{app:outliers} for
further details.

% However we are not able to determine in a 
%conclusive manner their physical cause.

As explained in Sec.~\ref{sec:intro}, the ultimate goal 
for wide parameter space searches for continuous signals over 
large data sets is to employ hierarchical schemes which alternate 
coherent and semi-coherent techniques. 
The Hough search would then be used  to 
select candidates in the parameter space to be 
followed up. The way in which those candidates would be selected is the 
following: fix the number of candidates to follow up, determine the false
 alarm rate, and set the corresponding threshold on the Hough number count. 
 Not all
the candidates selected in this manner will correspond to real gravitational
wave signals, but they will  point to interesting areas in parameter space.

The analysis presented here 
is a very important step forward in this direction. However, given the 
limited sensitivity its relevance mostly rests in the demonstration of 
this analysis technique on real data. In what follows we will thus 
concentrate on setting upper limits on the amplitude $h_0$ of continuous 
gravitational waves emitted at different frequencies.

%
%The way to rule out some of the strong candidates is to follow
%them up, for example, by performing a coherent search for a certain period of
%time. But it could happen that because of instrumental artifacts 
%there could be too many candidates making impossible to follow up them all.
%So there is a potential need to develop other vetoes. A suggested one is 
%to plot the loudest event for each sky position in a certain band, if 
%the plot shows that the loudest events are 90 degrees from the average
%acceleration vector of the Earth, i.e., a part of the sky with constant Doppler shift where instrument
%lines show up with the most power, then the candidate could be ruled out.
%Plots like this could help us to identify true candidates for
%follow up studies, but this requires further investigation.

%%%%%%%%%%%%%%%%%%%%%%%%%%%%%%%%%%%%%%%%%%%%%%%%%%%%%%%%%%%%%%%%%%%%%%%%%%%%%%%%%%%%%%%
\section{Upper limits}
\label{sec:ul}

%{\bf check. Our approach is perhaps not quite the standard frequentist approach},
We use a frequentist method to set upper limits on 
the amplitude $h_0$ of the gravitational wave signal. 
Our upper limits refer to a hypothetical population of isolated
spinning neutron stars which are uniformly distributed in the
sky and have a spin-down rate $\dot{f}$ uniformly distributed in the
range $(-\dot{f}_\max,0)$. We also assume uniform distributions for the
parameters $\cos\iota \in [-1,1]$, $\psi\in [0,2\pi]$, and $\Phi_0 \in
[0,2\pi]$.  As before, the frequency range considered is
$200$--$400\,$Hz.
%%%%%%%%%%%%%%%%%%%
%%We set upper limits on $h_0$ at different frequencies for a hypothetical
%%population of neutron stars uniformly distributed over the sky, and for which
%%the parameters $\cos\iota$, $\psi$, $\Phi_0$, $f_0$, and $\dot{f}$ are
%%also uniformly distributed within the parameter space region
%%under consideration.  
%%

The upper limits on %the gravitational wave strain 
$h_0$ emitted at
different frequencies are based on the highest number count, the {\it
loudest event}, registered over the entire sky and spin-down range
at that frequency. Furthermore, 
we choose to set upper limits not on each 
%the upper limits actually do not refer to a 
single frequency but on a set of frequency values lying
within the same $1\,$Hz band and thus are based on the highest number
count in each frequency band.  
In every $1\,$Hz band the loudest event % $n^\star$ 
is selected excluding all the vetoed frequencies of 
Table~\ref{tab:S2knownlines}. 

Let $n^\star$ be the loudest number count measured from the data.
The upper limit  $h_0^C$ on the gravitational wave amplitude, 
corresponding to a confidence level $C$, is  the value  such that
had there existed in the data a real signal with an amplitude greater 
than or equal to $h_0^C$, then in an ensemble of identical experiments 
with different  realizations of the noise, some fraction $C$ of trials  
would yield a maximum number count equal to or bigger than $n^\star$. 
%We will therefore say that we have placed an upper limit $h_0^C$
%on the gravitational wave strain, with confidence $C$.
The upper limit $h_0^C$ corresponding to a 
confidence level $C$ is thus defined as the solution to this equation: 
\begin{equation}\label{eq:uldef}
\textrm{Prob}(n \geq n^\star  |h_0^{C}) =
\sum_{n=n^\star}^{N} p(n|h_0^{C}) =  C\,,
\end{equation}
where $p(n|h_0)$
is the number count distribution in the presence of a signal with
amplitude $h_0$ and averaged over all the other parameters; note that
$p(n|h_0)$ is different from the distribution $p(n|h)$ discussed
before Eq.~(\ref{eq:binomialnosig}) which was relevant for a targeted
search when \emph{all} signal parameters are known.
We choose to set upper limits at a confidence level of $C=95\%$;
$h_0^{95\%}$ denotes the $95\%$ confidence upper limit value.

%%%%%%%%%%%%%%%%%%%%%%%%%
%%A series of software
%%injections of fake signals drawn from the population described above,
%%is then performed in the real data. The data with the injections is
%%searched using the search pipeline used for the actual analysis. This
%%is how the detectability of the population at a fixed strain amplitude
%%is measured. The strain amplitude is varied until the required
%%detection confidence is measured. Such value of the strain amplitude
%%is the upper limit.  This is described in Fig.
%%\ref{fig:MCpipeline}. 

%where $n^\star$ is the loudest number count measured from the data,
%$p(n|h_0)$ is the probability distribution for the number count $n$ in
%the presence of the target population of signals.  A confidence level
%of $C=95\%$ has been chosen; $h_0^{95\%}$ denotes the $95\%$
%confidence upper limit value.  

\begin{figure}
  \begin{center}
  \includegraphics[height=7cm]{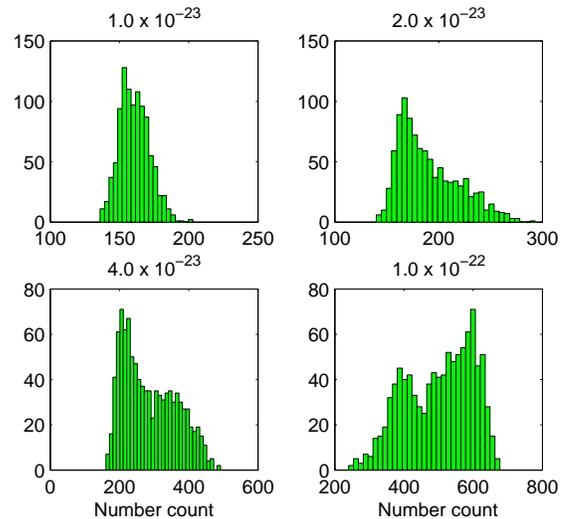}
  \caption{ Histograms of the Hough number count distribution (in arbitrary
  units) for L1
   using 1000 injected signals randomly distributed over the whole sky 
   within the band 200--$201\,$Hz
   for different  $h_0$ values.
   The largest number count for the search in that band was
   202. The  confidence level associated with the different $h_0$ values
   are: $0.1\%$ for $1.0 \times 10^{-23}$,  $30.5\%$ for $2.0 \times 10^{-23}$,
    $87.0\%$ for $4.0 \times 10^{-23}$, and $1$ for $1.0 \times 10^{-22}$.}
  \label{fig:L1histo4} 
  \end{center}
\end{figure}

\begin{figure}
  \begin{center}
  \includegraphics[height=9cm]{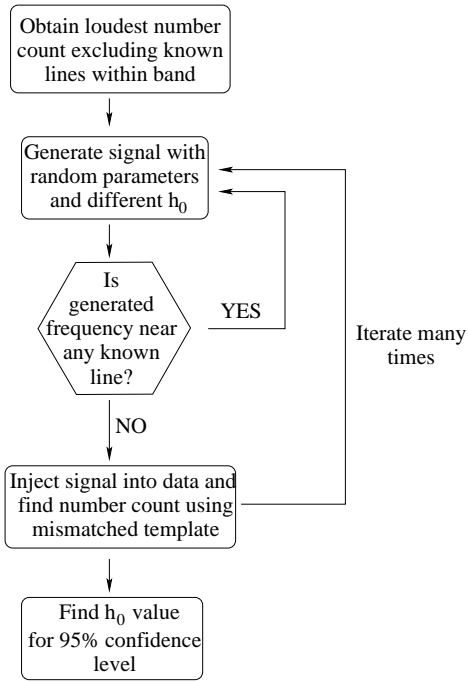}
  \caption{Pipeline description of the Monte-Carlo simulation to
  determine the upper limit values $h_0^{95\%}$.  We inject randomly
  generated fake pulsar signals with fixed amplitude (the other
  parameters are drawn from a suitable uniform distribution) into the
  real data and measure the value of $h_0$ required to reach a $95\%$
  confidence level. } 
  \label{fig:MCpipeline} 
  \end{center}
\end{figure}

Given the value of $n^\star$, if the distribution $p(n|h_0)$ were known,
it would be a simple matter to solve Eq.~(\ref{eq:uldef}) for
$h_0^C$.  In the absence of any signal, this distribution is indeed
just a binomial, and as exemplified in Fig.~\ref{fig:histo}, this is
also what is observed in practice. 
If a signal were present, the distribution may not be sufficiently 
close to binomial because the
quantity $\lambda$ defined in Eq.~(\ref{eq:lambda}) varies across the
SFTs due to non-stationarity in the noise and the amplitude modulation 
of the signal for different sky locations and pulsar orientations.   
In addition, now we must also consider the random mismatch between the
signal and template (in relation 
to the parameter space resolution used in the analysis) which causes an
additional reduction in the effective SNR for the template. 
%In addition, note that to obtain $p(n|h_0)$, we need to
%marginalize over the remaining signal parameters such as $\iota$, $\psi$ etc.  
%For these reasons, we measure $p(n|h_0)$ via Monte-Carlo simulations.  
%
For these reasons, we measure $p(n|h_0)$ by means of a series of 
software injections of fake signals in the real data. 
Fig.~\ref{fig:L1histo4} shows four
distributions for different $h_0$ values obtained by Monte Carlo
simulations. While for low signal amplitudes the distribution is close
to the ideal binomial one, the distribution diverges from it at higher
amplitudes, thereby illustrating the  complexity of the number count
distribution for sufficient large $h_0$.

Our strategy for
calculating the $95\%$ upper limits is to find $p(n|h_0)$ for a wide
range of $h_0$ values, then to get 
the corresponding confidence levels $C(h_0)$, and find the two values
of $h_0$ which enclose the $95\%$ confidence level.  The $95\%$ upper
limit is approximated by a linear interpolation between these values.  
We then refine this reduced range of $h_0$ values by further
Monte-Carlo simulations until the desired accuracy is reached.   
This is described in Fig.~\ref{fig:MCpipeline}.

The parameters of the fake injected signals are drawn from the population 
described above, and we ensure that the frequency does 
not lie within the excluded bands. 
The data with the injections are 
searched using the search pipeline used for the actual analysis. 
For computational efficiency, for each injected signal we find the
number count using only the 16 nearest templates,
 and choose the one yielding the largest number count. 
The SFT data in the different frequency bins in a $1\,$Hz band 
(of order $N\times 1800$ bins that get combined differently for the 
different time-frequency patterns) can be considered as different realizations 
of the same random process. Therefore it is reasonable to assume that the
normalized histogram of the largest number count represents the
discrete probability distribution $p(n|h_0)$. 

%The Monte-Carlo simulations are described in the flow-diagram in Figure
%\ref{fig:MCpipeline}.  For each $1\,$Hz band, the result of the search
%is a number count $n^\star$ which represents the loudest event in that
%frequency band, for the entire sky and the entire spindown range,
%which is not clearly due to a known noise artifact. Then we generate
%signals with 
%random parameters following the distributions mentioned above, 
%ensuring that the frequency does not lie within the excluded bands.
%The generated signals are injected into the noise SFTs.  Then we
%generate a template grid around this injected signal ensuring that
%this grid has the same resolution as the one used for the  search, i.e.
%the templates have a random mismatch from the signal. 
%For each injected signal we find the number count using all the 
%templates in this grid, %in particular we use only the 16 nearest ones,
%and choose the grid point yielding the largest number count.  
%This procedure is repeated a large number of times to obtain the
%distribution $p(n|h_0)$.  

The most computationally intensive part of this Monte-Carlo scheme is
the generation of the artificial signals.  The computational costs 
can be greatly reduced by estimating  $p(n|h_0)$ for different  $h_0$ values
in one go: for each individual artificial signal, we generate a set of
SFTs containing only one noiseless signal of a given amplitude. These
SFTs can be scaled by an appropriate numerical constant to obtain a
set of SFTs containing signals with different amplitudes, which are
then added to the noise SFTs.  The disadvantage of doing this is that
the different signals obtained by rescaling the amplitude this way are
not statistically independent since all the other signal parameters
are identical.  We must ensure that we have a sufficiently large
number of trials so that the error in the final upper limit is
sufficiently small.  

\begin{figure}
  \begin{center}
  \includegraphics[height=6cm]{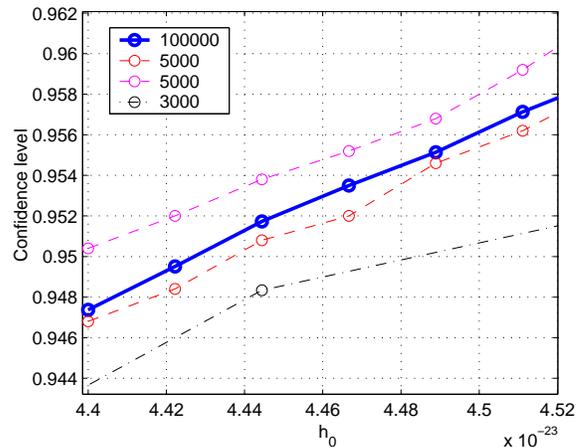}
  \caption{ Confidence level as a function of the signal amplitude $h_0$
  for different Monte Carlo simulations for L1 within the band 200--$201\,$Hz. 
  The solid  line corresponds
  to a simulation using 100000 injected signals. The two 
  dashed lines correspond to 5000 injected signals. The dash-dotted  line
  corresponds to 3000 signal injections. 
  The $h_0^{95\%}$   upper limit
   for this band using and these two simulations with
   5000 injections has a maximum absolute error of the order of
   $0.02 \times 10^{-23}$ corresponding to a relative error smaller 
   than  $0.5\%$.  In the case of using only 3000 injections the error increases
   to the $1.5\%$ level.}
  \label{fig:L1_200_Ch0} 
  \end{center}
\end{figure}

\begin{figure}
  \begin{center}
  \includegraphics[height=6cm]{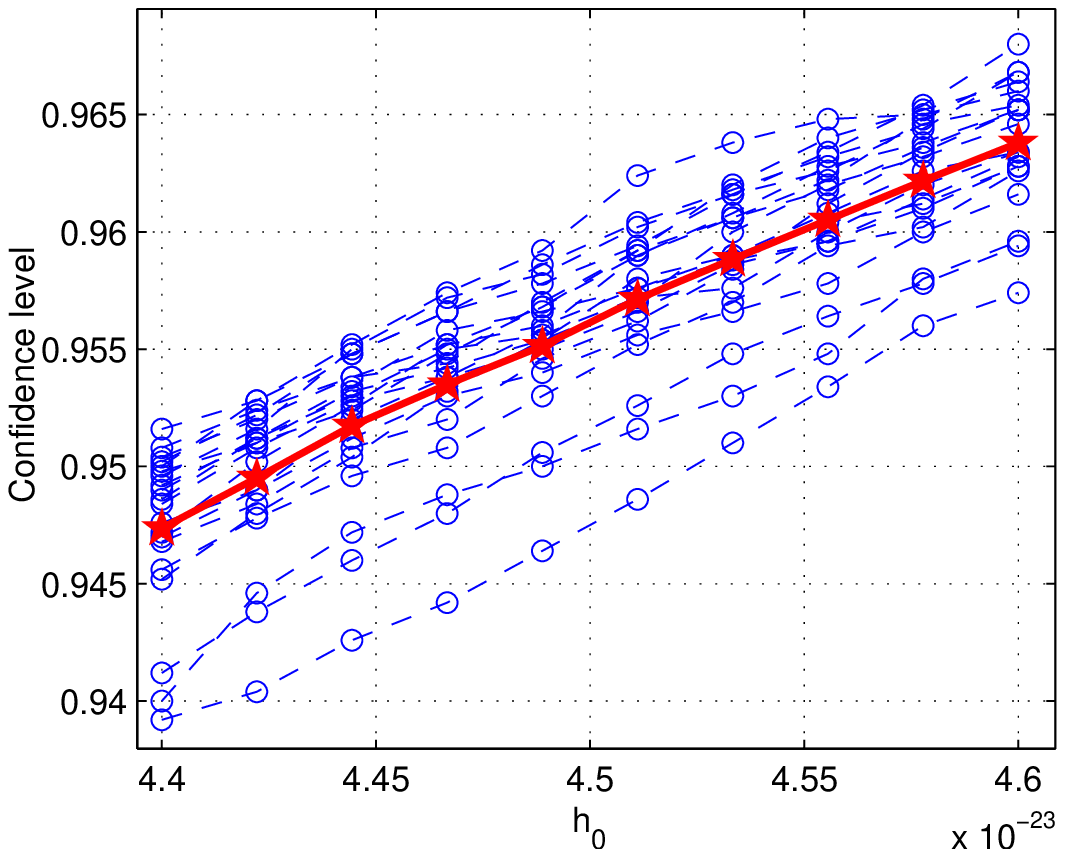}
  \caption{ Confidence level as a function of the signal amplitude $h_0$
  for different Monte Carlo simulations for L1 within the band $200$--$201\,$Hz. 
  The solid thick line corresponds
  to a simulation using 100000 injected signals. The other 20 lines
  correspond to simulations with 5000 injected signals. 
  The $h_0^{95\%}$ upper limit
   for this band using 5000 injections has  a maximum absolute error of 
   the order of
   $0.1 \times 10^{-23}$ corresponding to a relative error of
   $2.2\%$. }
  \label{fig:L1_200x20} 
  \end{center}
\end{figure}

\begin{figure}
  \begin{center}
  \includegraphics[height=6cm]{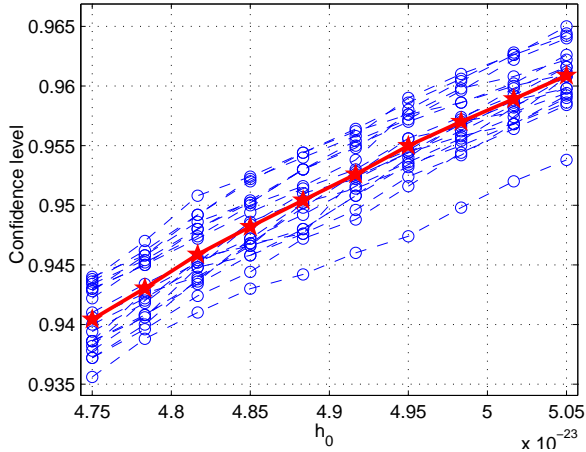}
  \caption{ Confidence level as a function of the signal amplitude $h_0$
  for different Monte Carlo simulations for H1 within the band $259$--$260\,$Hz. 
  The solid thick  line corresponds
  to a simulation using 100000 injected signals. The other 20 lines
  correspond to simulations with 5000 injected signals. 
  The $h_0^{95\%}$ upper limit
   for this band using 5000 injections has a maximum absolute error of
    the order of
   $0.1 \times 10^{-23}$ corresponding to a relative error of
   $2.0\%$. }
  \label{fig:H1259chx20} 
  \end{center}
\end{figure}
We have found empirically that 5000 injections per band are sufficient
to get upper limits accurate to within $3\%$; see 
Fig.~\ref{fig:L1_200_Ch0}, \ref{fig:L1_200x20} and \ref{fig:H1259chx20}.  
Fig.~\ref{fig:L1_200_Ch0} shows the confidence level as a function of the 
signal amplitude $h_0$ for L1 data within the band $200$--$201\,$Hz.  
The solid line corresponds to our most accurate simulations using
100,000 injections. The two dashed lines correspond to two different
simulations both using 5000 injected signals and  the dotted line 
corresponds to simulation with 3000 injections. In each case,
confidence levels for different $h_0$ values are calculated by
simply rescaling the signal as described above.  This means that all
the points in each individual curve in Fig.~\ref{fig:L1_200_Ch0} are
statistically biased in a similar way, and this explains why 
the curves in this figure do not intersect.  To estimate the error in the
$95\%$ upper limit, we generate several of these curves for a fixed
number of injections and we measure the error in $h_0^{95\%}$ 
with respect to the accurate reference value obtained
using 100,000 injections.  For this particular band and detector, we
find that for 5000 injections, the error in the upper limit is at most
$0.1 \times 10^{-23}$, corresponding to a relative error of
$2.2\%$. This experiment has been repeated for several $1\,$Hz bands and
many simulations, and in all of them 5000 injections 
were enough to ensure an accuracy of better than $3 \%$ in the
$h_0^{95\%}$. We also have found that with 5000 injections, 
using amplitudes equal to the most accurate upper limit $h_0^{95\%}$
obtained from 100,000 injections, the confidence level 
are within $94.5-95.5 \%$. See Fig.~\ref{fig:L1_200x20}
and \ref{fig:H1259chx20}.

\begin{figure} 
  \begin{center} 
  (a) \\
  \includegraphics[height=3.5cm]{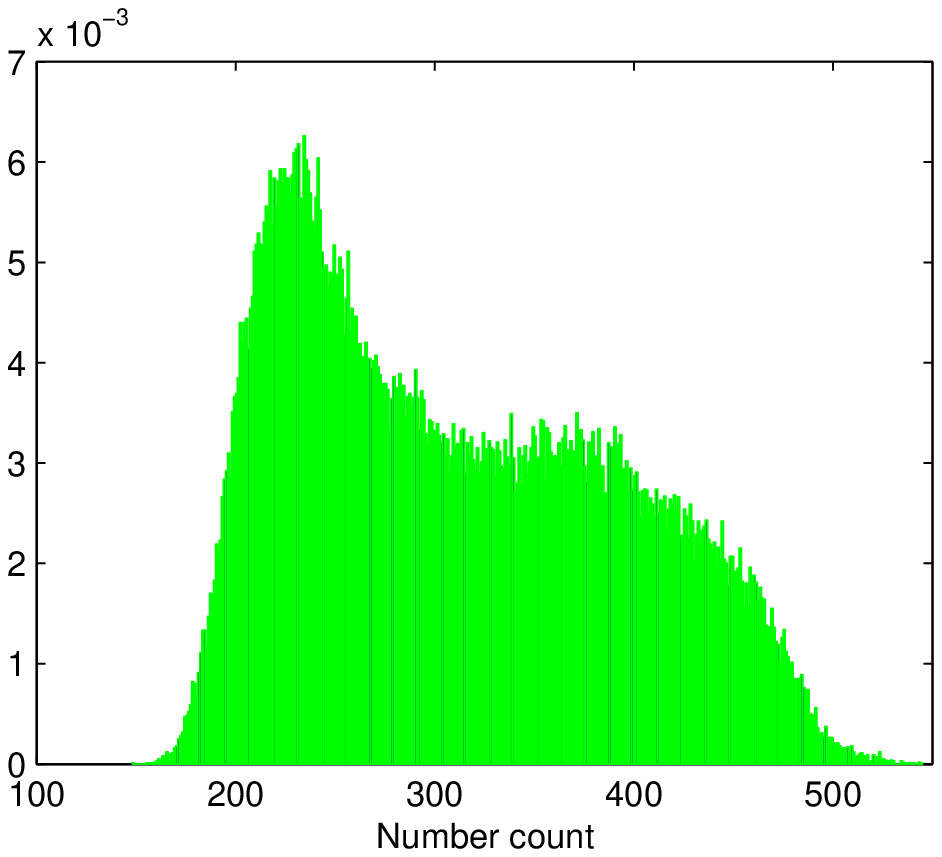}\\
  \begin{tabular}{cc}
  (b) & (c) \\
   \includegraphics[height=3.5cm]{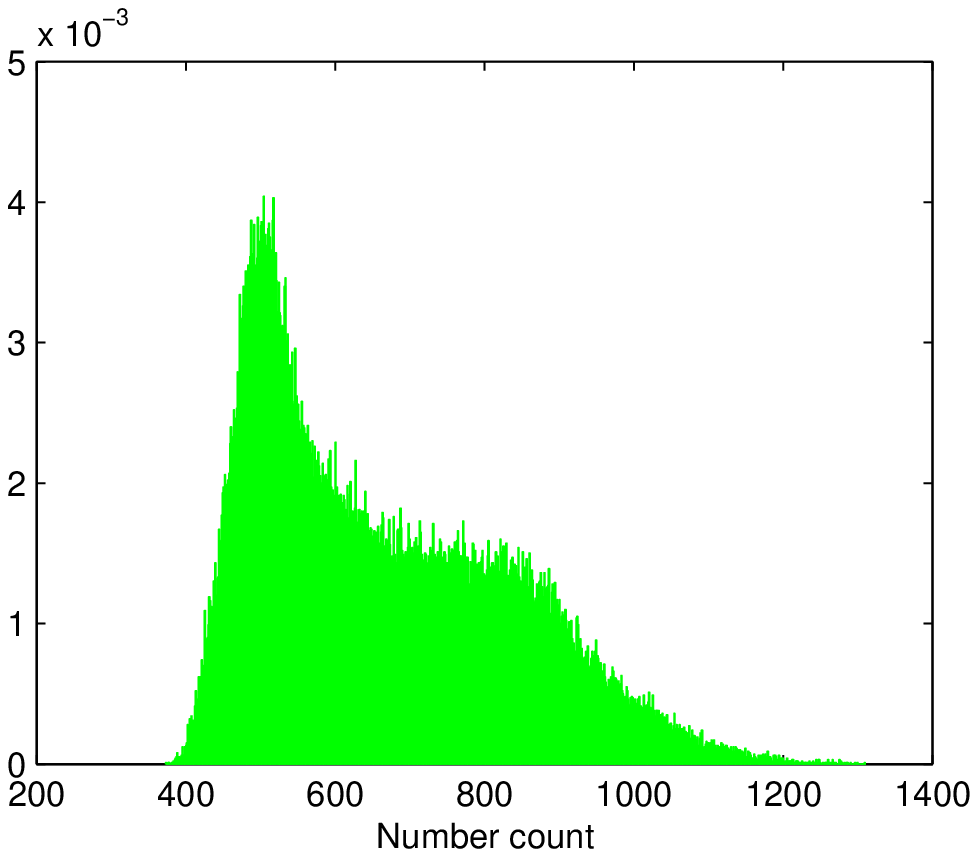} &
   \includegraphics[height=3.5cm]{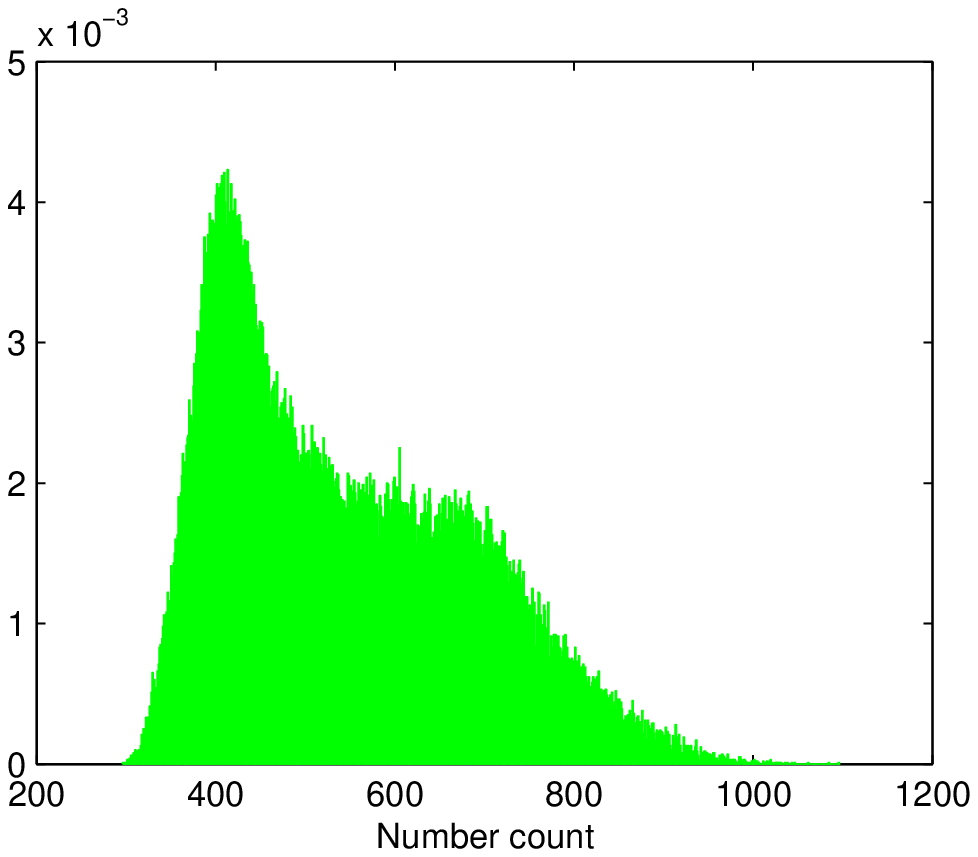}
   \end{tabular}
  \caption{Measured Hough number count probability distribution for 
  $p(n|h_0^{\inject})$ from the Monte Carlo 
  simulations using 100,000 injected signals. (a) L1 within the 
  band 200--$201\,$Hz, with a $h_0^{\inject}$  of 
  $4.422 \times 10^{-23}$ corresponding to a confidence level of $94.95\%$.
   The $n^\star$ value for this band was 202. 
   (b) H1 within the band 259--$260\,$Hz, with a $h_0^{\inject}$ of 
  $4.883 \times 10^{-23}$ corresponding to a confidence level of $95.04\%$. 
  The $n^\star$ value for this band was 455.
  (c) H2 within the band 258--$259\,$Hz, with a $h_0^{\inject}$  of 
  $8.328 \times 10^{-23}$ corresponding to a confidence level of $95.02\%$. 
  The $n^\star$ value for this band was 367.}
  \label{fig:hist} 
  \end{center}
\end{figure}

Fig.~\ref{fig:hist}  shows the
probability distribution  for $p(n|h_0^{\inject})$ for L1, H1 and H2 in
a $1\,$Hz band measured from
100,000 randomly injected signals, with a signal amplitude $h_0^{\inject}$
\emph{close} to the $95\%$ confidence level; note that these
$h_0^{\inject}$ values are not the $95\%$ upper limits because they
correspond to different confidence levels. This illustrates, yet
again, that the true number count distributions are 
far from the ideal binomial distribution.
%The $h_0^{\inject}$  value used 
%is $4.422 \times 10^{-23}$ corresponding to the $94.95\%$ confident upper
%limit. 
%Figure \ref{fig:H1259hist}  shows the pdf distribution  
%for H1 within the band 259-260 Hz. 
%The $h_0^{\inject}$  value used 
%is $4.883 \times 10^{-23}$ corresponding to a $95.04\%$ confident upper
%limit. 
%Figure \ref{fig:H2258hist}  shows the pdf distribution  
%for H2 within the band 258-259 Hz. 
%The $h_0^{\inject}$  value used 
%is $8.328 \times 10^{-23}$ corresponding to a $95.02\%$ confident upper
%limit.
It is also interesting to notice that, among all the
templates used to analyze each injection,
the nearest template, in the normal Euclidean sense, corresponds to 
the template providing the highest number count only $18\%$ of the time.
This effect is due  to the noise contribution and
the fact that the matching-amplitude, described by the parameter space
metric is highly anisotropic \cite{prix05}. 

%{\bf can we say how far the number count at the nearest template wrt
%to the highest number count ? Perhaps not for the paper.} 
%
%It is worth mention that for these Monte Carlo signal injection, among all the
%templates used to analyze each injection (typically the 16 nearest templates),
%the nearest template, in the normal Euclidean sense, corresponded to 
%the template
%providing the highest number count only in $18\%$ of the times.
%This is not surprising and it is due to two reasons: i)
%the noise contribution  and ii)
%the fact that the matching-amplitude, 
%described by the {\it metric} is highly anisotropic. Being closer in the normal
%Euclidean way does not guarantee having the highest signal-amplitude, and
%this is independent of the grid-size. A point lying along the
%main-axis of the n-dimensional ellipse can have a higher number 
%count even if it is further away in the Euclidean sense.
%This effect is well known for the  coherent searches.
% Assuming the incoherent Hough metric to be a kind of average
%over these 30 minutes metrics, and since the ellipses change 
%orientation and shape  over the whole sky as a function of time,
%one would indeed expect them to become more isotropic than for the 
%fully coherent search.

\begin{figure} 
  \begin{center}
  \includegraphics[height=6.5cm]{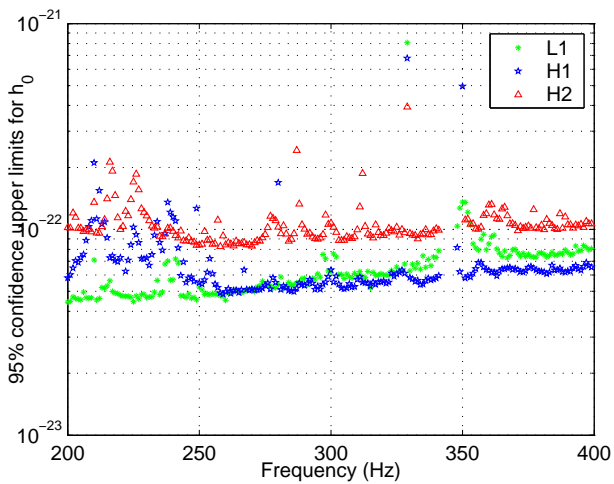}
  \caption{The $95\%$ confidence  upper limits on $h_0$ over the whole sky
  and different spin-down values in $1\,$Hz bands.
  }
  \label{fig:ul95} 
  \end{center}
\end{figure}

The $95\%$ confidence upper limits on $h_0$ for each $1\,$Hz
band using all the data from the S2 run are shown in 
Fig.~\ref{fig:ul95}.  As expected, the results are very similar for the L1
and H1 interferometers, but significantly worse for H2.
The typical upper limits in this frequency range for
L1 and H1 are mostly between $\sim 4$--$9\times 10^{-23}$, typically
better at lower frequencies. The most stringent upper limit
for L1 is $4.43\times
10^{-23}$ which occurs within $200$--$201\,$Hz, largely reflecting 
the lower noise floor around $200\,$Hz.
For H1 it is $4.88\times 10^{-23}$, which occurs in the frequency range
$259$--$260\,$Hz, and for H2 it is $8.32\times
10^{-23}$, which occurs in the frequency range 
$258$--$259\,$Hz. The values of the most stringent upper limits on 
$h_0^{95\%}$ have been 
obtained using 100,000  injections in the most sensitive $1\,$Hz bands.
The upper limits are significantly worse in bands lying near the known
spectral disturbances, especially near the violin modes.

In Table~\ref{tab:S2ulsum} we summarize the best upper limits on $h_0^{95\%}$
and we compare them with the  theoretical values $h_0^{\Exp}$
we would expect for a directed search using a perfectly matched template, 
as given by Eq.~(\ref{eq:sensitivity}). 
Here we take a false dismissal rate of $5\%$
and the false alarm rate associated to the loudest number count in that band.
In those three bands the ratio   $h_0^{95\%}/h_0^{\Exp}$ is about 1.8.

These $h_0^{95\%}$ results are also about a factor of 2.6 worse than 
those  predicted by Fig.~\ref{fig:S2sensitivity} %Eq.~(\ref{eq:sensitivity1-10}) 
which corresponds to a directed
search using a perfectly matched template, and with a false alarm rate
of $1\%$ and a false dismissal rate of $10\%$.  Of these 2.6, a factor 
$\sim 1.5$ is due to the use of different values of the false alarm 
and false dismissal
rate, and a factor $\sim 1.8$ because the $p(n|h_0)$ distribution does not 
correspond to the ideal binomial one for values of $h_0$ distinct from zero.

\begin{figure}
  \begin{center}
  \includegraphics[height=6.5cm]{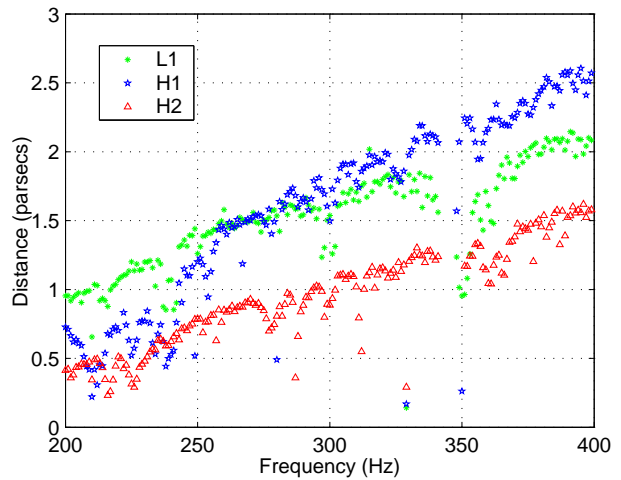}
  \caption{ Astrophysical reach covered by the search. The
  curves show the estimated distance $d$ out to which signals
  from isolated gravitational wave pulsars could be detected in our
  S2 data set derived from the upper limits on $h_0$. This plot 
  assumes $\epsilon = 10^{-6}$ and $I_{zz}=10^{45}$g\,cm$^2$. }
  \label{fig:S2reachUL} 
  \end{center}
\end{figure}

From the upper limits on $h_0$, using Eq.~(\ref{eq:h0}), we can derive
the distance covered by this search. This is shown in
Fig.~\ref{fig:S2reachUL} assuming $\epsilon = 10^{-6}$ and
$I_{zz}=10^{45}$g\,cm$^2$. The maximum reach is 2.60 pc for H1 which
occurs within $395$--$396\,$Hz. For L1 the maximum reach is 2.15 pc and
1.62 pc for H2.  This value of $\epsilon=10^{-6}$ corresponds to  the
maximum expected ellipticity for a 
regular neutron star, but ellipticities from more exotic
alternatives to neutron stars may be larger~\cite{Owen:2005fn}, {\it e.g.},
solid strange quark stars for which $\epsilon_{\max} \approx10^{-4}$.
Therefore the astrophysical reach for these exotic
stars could be better by a factor 100.

\begin{widetext}
\begin{center}
\begin{table}
\begin{tabular}{cccccccccc}\hline
Detector & Best  $h_0^{95\%}$ & $f_{\band}$ (Hz) & $\langle S_n(f_{\band})
\rangle$ (Hz$^{-1}$) & $N$ & $n^*$ & $\alpha^*$ &  $m_{1Hz}$ &
$\mathcal{S}^{\Exp}$ & $h_0^{\Exp}$ \\ \hline \hline
L1 & $4.43\times 10^{-23}$ & 200-201 & $1.77\times 10^{-43}$ & 687  & 202 &
$8.94\times 10^{-10}$ & $1.88\times 10^{9}$ &5.4171 & $2.41\times 10^{-23}$\\
H1 & $4.88\times 10^{-23}$ & 259-260 & $3.53\times 10^{-43}$ & 1761 & 455 &
$1.77\times 10^{-9}$ & $3.11\times 10^{9}$ & 5.3381 & $2.67\times 10^{-23}$\\
H2 & $8.32\times 10^{-23}$ & 258-259 & $1.01\times 10^{-42}$ & 1384 & 367 &
$2.25\times 10^{-9}$ & $3.11\times 10^{9}$& 5.3098 & $4.78\times 10^{-23}$\\
\hline
\end{tabular}
\caption{ Best all-sky upper limits obtained on the strength of 
gravitational waves from isolated neutron stars.
The $h_0^{95\%}$ values have been obtained using 100,000 
injections in the best $1\,$Hz bands $f_{\band}$ for the three detectors.
$\langle S_n(f_{\band})\rangle$ is the average value of noise in that $1\,$Hz band
excluding the vetoed frequencies, 
$N$ the number of SFTs available for the
entire S2 run,  
$n^*$ is the loudest number-count measured from the data in that
band,  
$\alpha^*$ is the corresponding false alarm assuming Gaussian
stationary noise derived from Eq.~(\ref{eq:nth}), 
 $m_{1\,Hz}$ the number of templates analyzed in that $1\,$Hz band,
the quantify $\mathcal{S}^{\Exp}$ is defined by  Eq.~(\ref{eq:S}) using
 the values $\alpha=\alpha^*$ and $\beta=0.05$.
$h_0^{\Exp}$ is the theoretical expected upper limit from searches 
with perfectly matched templates assuming the ideal binomial distribution
for $p(n|h_0)$ defined by Eq.~(\ref{eq:sensitivity}), therefore ignoring
also the effects of the different sensitivity at different sky locations
and pulsar orientations.
} 
\label{tab:S2ulsum}
\end{table}
\end{center}
\end{widetext}
%

%
%\begin{figure}
%  \begin{center}
%  \includegraphics[height=6.5cm]{ellipticity}
%  \caption{ Upper limits on ellipticity $\epsilon$ up to a distance of 2.60 pc
%   and $I_{zz}=10^{45}$g\,cm$^2$. }
%  \label{fig:S2ellipticity} 
%  \end{center}
%\end{figure}
%

% 
%For a given distance, the upper limits can also be interpreted as 
%upper limits on the ellipticity. This is shown in Fig.~\ref{fig:S2ellipticity} 
%assuming the maximum distance of 2.60 pc obtained before.

%%%%%%%%%%%%%%%%%%%%%%%%%%%%%%%%%%%%%%%%%%%%%%%%%%%%%%%%%%%%%%%%%%%%%%%%%%%%%%%%%%%%%%%%%%%%
\section{Hardware injections}
\label{sec:hardwareinj}

Two artificial pulsar signals, based on the waveforms given in
Eqs.~(\ref{eq:detoutput}), (\ref{eq:waveform1}), (\ref{eq:waveform2}),
 and (\ref{eq:phasemodel}) were injected into all three
LIGO interferometers for a 
period of 12 hours towards the end of the S2 run.  These injections were
designed to give an end-to-end validation of the search pipeline.  The
waveforms were added digitally into the interferometer Length Sensing
and Control System (responsible for maintaining a given interferometer
on resonance), resulting in a differential length dither in the
optical cavities of the detector. We denote the two pulsars P1
and P2; their parameters are given in Table~\ref{tab:S2hdw}.   
\begin{table}
\begin{tabular}{lcc}\hline
       & P1 & P2 \\ 
\hline \hline
$f_0$ (Hz)  & 1279.123456789012 & 1288.901234567890123 \\
$\dot{f}$ (Hz-s$^{-1}$) & 0 & $-10^{-8}$ \\
RA (rad) &  5.1471621319 & 2.345678901234567890 \\
Dec (rad) & 0.3766960246 & 1.23456789012345 \\
$\psi$ (rad) & 0 & 0    \\
$\cos\iota$ & 0 & 0 \\
$\Phi_0$ (rad) & 0 & 0 \\
$T_0$ (sec) & 733967667.026112310 & 733967751.522490380\\
$h_0$ & $2 \times 10^{-21}$ & $2\times 10^{-21}$ \\
\hline
\end{tabular}
\caption{ Parameters of the two hardware injected pulsars.  See
  Eqs.~(\ref{eq:detoutput}), (\ref{eq:waveform1}),
  (\ref{eq:waveform2}),  and (\ref{eq:phasemodel}) for
  the definition of the parameters. RA and Dec are the right ascension 
   and declination in equatorial coordinates.
  $T_0$ is the GPS time 
  in the SSB frame in which the signal parameters are defined. 
  $h_0$ is the amplitude of the signal according to
  the  strain calibration used at the time of the injections.
   }\label{tab:S2hdw}
\end{table}
\begin{figure}
  \begin{center}
  \includegraphics[height=11cm]{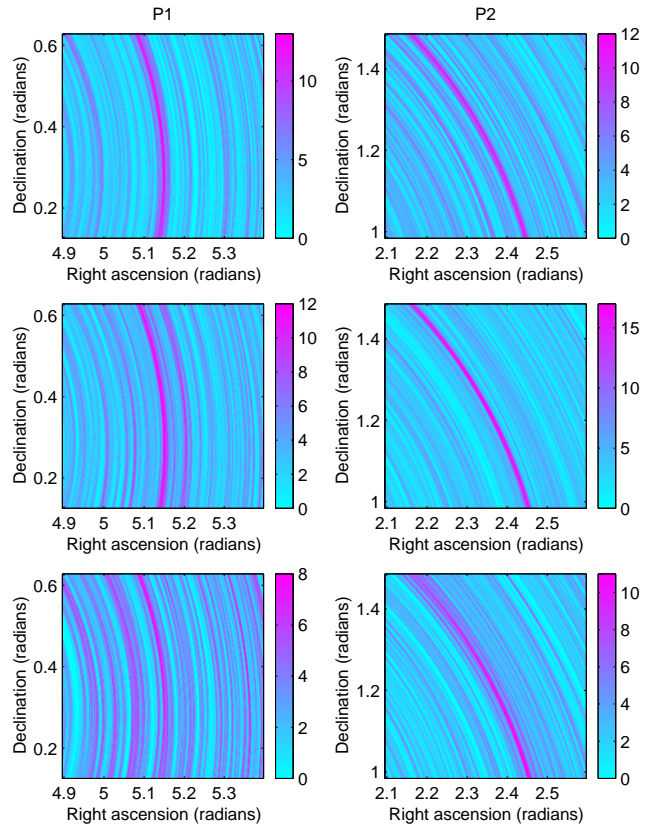}
  \caption{Hough maps for the hardware injected signals. The left maps
  correspond to P1 and the right maps to P2. From top to bottom, the left maps
  correspond to  L1, H1 and H2, for a frequency of $1279.123333\,$Hz and zero 
  spin-down; the right maps to a frequency of  $1288.901111\,$Hz and zero, 
  $-1.77024\times 10^{-8}\,$Hz-s$^{-1}$, and
   $-1.93533\times 10^{-8}\,$Hz-s$^{-1}$ spin-down values respectively.
   The location of the injected signals are close to the centers of these
   subplots.}
  \label{fig:S2injections}
  \end{center}
\end{figure}

The data corresponding to the injection period have been analyzed using the
Hough transform and the same search code as described in 
Sec.~\ref{subsec:pipeline}. 
As before, the input data consists of 30 minutes long
SFTs. The number of SFTs available are 14 for L1, 17 for H1 and 13 for
H2.  As in Sec. \ref{subsec:parameters}, the frequency resolution
is $1/1800\,\textrm{s}$, the sky resolution is given by equation
(\ref{eq:minannulus}), and $\delta\dot{f} = 1/(\Tobs\Tcoh)$.   Since
the total effective observation time is somewhat 
different for the three detectors, we get $\delta\dot{f} =
-2.28624\times 10^{-8}\,$Hz-s$^{-1}$ for L1, $-1.77024\times
10^{-8}\,$Hz-s$^{-1}$ for H1, and $-1.93533\times
10^{-8}\,$Hz-s$^{-1}$ for  
H2. As before, for each intrinsic frequency we analyze 11 different
spin-down values. The portion of sky analyzed was of $0.5$ radians
$\times$ $0.5$ radians around the location of the two injected
signals. 

Fig.~\ref{fig:S2injections} shows the Hough maps corresponding to the 
nearest frequency and spin-down values to the injected ones. Although 
the presence of the signal is clearly visible, it is apparent that 12 hours 
of integration time is not enough to identify the location of the source 
using the Hough transform. In particular if one looks
at the Hough maps at mismatched frequencies and spin-down values, one
can still identify annuli with very high number-counts, but appearing
with a mismatched sky location.  

For P2, the Hough maps
corresponding to the closest values of frequency and spin-down
contain the maximum number count at the correct sky location.
These maximum number counts are  12 for L1, 17 for H1, and 11 for
H2. Notice that for L1 and H2 these numbers are smaller than the
number of SFTs used.  

For P1, the  closest template to the signal parameters has a number
count of 13 for L1, 12 for H1 and 8 for H2.  The maximum number
counts obtained in the search were $13$ for L1, but 14 for H1, and 10
for 
H2. Those higher number counts occurred for several templates with a
larger mismatch, for example at
$1279.123333\,$Hz and  $-5.31073\times 10^{-8}\,$Hz-s$^{-1}$ for H1, and 
$1279.132222\,$Hz and $-1.35473\times 10^{-7}\,$Hz-s$^{-1}$ for
H2. This is not surprising because we only compute the Hough maps at
the Fourier frequencies $n\times 1/\Tcoh$.  
%The injected frequency of
%P1 lies exactly in between the two Fourier frequencies
%1279.123333 Hz and 1279.123889 Hz, having therefore the largest possible
%frequency mismatch.  
In any case, both pulsar signals are
unambiguously detected because these number counts are much bigger
then the expected average number counts for pure noise.

%%%%%%%%%%%%%%%%%%%%%%%%%%%%%%%%%%%%%%%%%%%%%%%%%%%%%%%%%%%%%%%%%%%%%%%%%%%%%%%%%%%%%%%%%%%%%%
\section{Conclusions}
\label{sec:conc}

In this paper we use the Hough transform to search for periodic
gravitational-wave signals. This is a semi-coherent 
sub-optimal method.  Its virtues are computational efficiency and
robustness.  
The search pipeline was validated using a series of unit
tests and comparisons with independently written code.
We then applied this method to analyze data from the second
science run of all three LIGO interferometers.
We also validated the search pipeline by analyzing data from
times when two artificial pulsar signals were physically injected
into the interferometer hardware itself. We show
in this paper that the injected signals were clearly detected. 

\begin{table}
\begin{tabular}{ccc}\hline
        
Detector &  $f_{\band}$ (Hz)   & $h_0^{95\%}$ \\ 
\hline \hline
L1 & 200-201 & $4.43\times 10^{-23}$ \\
H1 & 259-260 &  $4.88\times 10^{-23}$ \\
H2 & 258-259 &  $8.32\times 10^{-23}$ \\
\hline
\end{tabular}
\caption{ Best all-sky upper limits obtained on the strength of 
gravitational waves from isolated neutron stars.
The $h_0^{95\%}$ values have been obtained using 100,000 
injections in the best 1\,Hz bands.} 
\label{tab:S2ul}
\end{table}

Our final results are all-sky upper limits on the
gravitational wave amplitude for a set of frequency bands. The best
upper limits that we obtained for the three interferometers are
given in Table~\ref{tab:S2ul}.  The overall best upper
limit is $4.43 \times 10^{-23}$. 
We searched the $200-400 \, \textrm{Hz}$ band and the spin-down space 
$\dot f_\max \le 1.1\times 10^{-9}\,\textrm{Hz-s}^{-1}$, and no vetoes were
applied except for the list of ignored frequency bands that contain instrument
line artifacts.  Our best upper limit is
26 times worse than the best upper limit 
obtained  for a \emph{targeted} coherent search using
the same data. This was an upper limit of $1.7\times 10^{-24}$ \cite{cw-prl},
achieved for PSR J1910-5959D.  This is to be expected because we have
performed not a targeted, but a wide parameter space search.  If we
were to use the optimal $\F$-statistic method to perform a
hypothetical search over the same
parameter space region as the Hough search in this paper, the number
of templates required would be much larger for the same observation
time: $\sim 10^{19}$ \cite{BCCS} instead of $\sim 10^{12}$.  Thus,
we would have to set a lower false alarm rate for this hypothetical
search, and in the end, the sensitivity turns out to be roughly
comparable to that of the Hough search.  Note also that this
hypothetical search is not computationally feasible for the
foreseeable future.

We can search for smaller signals either by increasing the number of
coherent segments, $N$, or by increasing the
coherent time baseline, $\Tcoh$.  Since the number of segments is 
determined by the length of the data set, for a given amount of data
one wants to make $\Tcoh$ as large as possible. However, for the search 
pipeline presented in this paper, increasing $\Tcoh$ was not possible
due to the restriction on its value mentioned in Sec. \ref{subsec:houghbasics}.
This will be overcome by demodulating each short segment, taking
into account the frequency and amplitude modulation of the signal.
The optimal method will be to calculate the $\F$-statistic \cite{jks} for
each segment. The time-frequency pattern will then no longer be given
by (\ref{eq:master}) and (\ref{eq:fhat}) but by the master equation
given in reference \cite{hough04}.  

Since the wide parameter-space
search for periodic gravitational-wave signals is computationally
limited, there is also a limit on the maximum $\Tcoh$ that can be
used, given finite computing resources. Thus, a hierarchical strategy
that combines fully-coherent and semi-coherent methods
will be needed to achieve optimal results \cite{BC00,cgk}. 
Our goal is to use the Hough transform as part of such a strategy.
This is work in progress and the results will be presented elsewhere.

%%%%%%%%%%%%%%%%%%%%%%%%%%%%%%%%%%%%%%%%%%%%%%%%%%%%%%%%%%%%%%%%%%%%%%%%%%%%%%%%%%%%%%%%%%%%%%%
\section*{Acknowledgments}

The authors gratefully acknowledge the support of the United States National Science
 Foundation for the construction and operation of the LIGO Laboratory and the 
 Particle Physics and Astronomy Research Council of the United Kingdom, the Max-Planck-Society 
 and the State of Niedersachsen/Germany for support of the construction and 
 operation of the GEO600 detector. The authors also gratefully acknowledge the support
  of the research by these agencies and by the Australian Research Council, the 
  Natural Sciences and Engineering Research Council of Canada, the Council of Scientific
 and Industrial Research of India, the Department of Science and Technology of India, 
 the Spanish Ministerio de Educaci\'on y Ciencia, the John Simon Guggenheim Foundation, 
 the David and Lucile Packard Foundation, the Research Corporation, 
 and the Alfred P. Sloan Foundation.
This document  has been assigned LIGO Laboratory document number
LIGO-P050013-00-R.

%%%%%%%%%%%%%%%%%%%%%%%%%%%%%%%%%%%%%%%%%%%%%%%%%%%%%%%%%%%%%%%%%%%%%%%%%%%%%%%%%%%%%%%%%%%
\begin{appendix}

\section{The bias in the running median}
\label{sec:bias}

We are using a running median to estimate the noise floor in our
SFTs.  Thus, the value of $S_n$ at a particular frequency bin can be estimated
from the
\emph{median} of $|\tilde{x}_k|^2$ in $w$ frequency bins around
the frequency bin, where $w$ is an integer and represents the
window-size of the running median. 
The reason for using a running median is to minimize the effect of
large spectral disturbances, and to ensure that the
presence of any pulsar signals will not bias our estimation of $S_n$.
To carry this out in practice, we would like to know how the median
can be used as an estimator of the mean.  In this appendix, we answer
this question assuming that the noise is Gaussian, so that the power
is distributed exponentially.   
\begin{figure}
  \begin{center}
  \includegraphics[height=6cm]{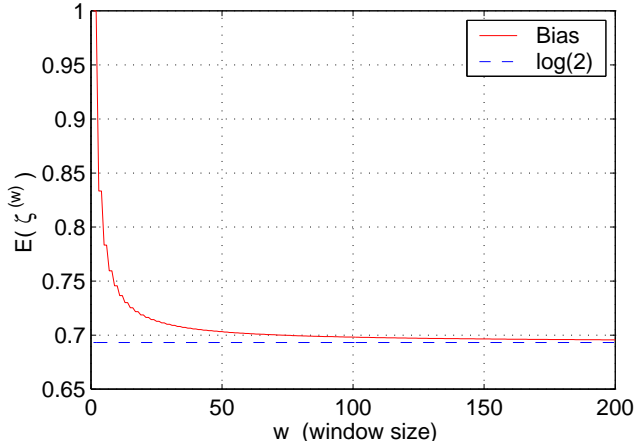}
  \caption{Value of the running median bias $\zeta^{(w)}$ as a
  function of the window size $w$; $\zeta^{(w)}$ approaches
  $\log(2.0)$ for large $w$.  }
  \label{fig:bias}
  \end{center}
\end{figure}

Let $x$ be a random variable with probability distribution $f(x)$.
Let $F(x)$ denote the cumulative distribution function:
\begin{equation}
F(x) = \int_{-\infty}^x f(x^\prime) dx^\prime\,.
\end{equation}
Let us draw $w$ samples from this distribution and arrange them in
increasing order: ${x_n}$ ($n=1\ldots w$).  Define an integer $k$ which is
$(w-1)/2$ when $w$ is odd, and $w/2$ when $w$ is even.  We define the
median $\zeta^{(w)}$ as 
\begin{equation}
\zeta^{(w)} = \left\{  \begin{array}{cr}
                        x_{k+1} & \textrm{when $w$ is odd\,,} \\
			\frac{1}{2}(x_k + x_{k+1}) & \textrm{when $w$
                        is even\,.}
    		      \end{array}
    \right.
\end{equation}
Consider first the case when $w$ is odd. The distribution of
$\zeta^{(w)}$ can be found as follows: $\zeta^{(w)}$ lies
within the range $(x,x+dx)$ when $k$ values are less than $x$, $w-k-1$
values are greater than $x+ dx$, and one value is in the range
$(x,x+dx)$. The probability density for $\zeta^{(w)}$ is thus 
\begin{equation} \label{eq:oddbias}
g_w(x) = \left(\begin{array}{cc}w\\k\end{array}\right) (w-k) [F(x)]^k
  [1-F(x)]^{w-k-1} f(x)\,.
\end{equation}
When $w$ is very large, it can be shown that the distribution $g_w(x)$
approaches a Gaussian whose mean is equal to the population median
$\hat{\zeta}$, and whose variance is proportional to $1/\sqrt{w}$
\cite{kramer}.  

For Gaussian noise, the normalized power $\rho$ follows the
exponential distribution, i.e. $f(x)= e^{-x}$ for $x\geq
0$ and $f(x)=0$ for $x<0$.  The mean is unity, therefore $\zeta^{(w)}/
\mathbf{E}(\zeta^{(w)})$ is an unbiased estimator of the mean, where
$\mathbf{E}(\zeta^{(w)})$ is the expectation value of 
$\zeta^{(w)}$: 
\begin{equation}
\mathbf{E}(\zeta^{(w)}) = \int_{-\infty}^{\infty}xg_w(x)dx  \,.
\end{equation}
We can explicitly calculate $\mathbf{E}(\zeta^{(w)})$ for this
case and the answer turns out to be given by a truncated alternating
harmonic series:  
\begin{equation}
\mathbf{E}(\zeta^{(w)}) = \sum_{j=1}^w\frac{(-1)^{j+1}}{j}\,.
\end{equation}
For very large $w$, $\mathbf{E}(\zeta^{(w)})$  approaches
$\ln(2)$, which is precisely the population median. For $w=1$ it
is just unity, which makes sense because in this case the median is
equal to the mean.    
%The window size $w$ should be as small as possible in order to track the
%noise floor accurately.  On the other hand, the statistical errors
%in the value of the running median are smaller when a large window 
%size is chosen. We have found that $w=101$ is a good compromise.
For finite $w$, $\mathbf{E}(\zeta^{(w)})$ is somewhat larger than
$\ln(2)$ and for a window size of $w=101$, which  
is what is used in the actual search, it is $ 0.698073$.  This
is to be compared with $\ln(2) = 0.693147$, a difference of
about $0.7\%$.  

When $w$ is even, the distribution of $\zeta^{(w)}$ is given (up to
a factor of $2$) by the convolution of the distributions of $x_k$ and
$x_{k+1}$.  However, we are interested only in the expectation value
\begin{equation}
\mathbf{E}(\zeta^{(w)}) = \frac{1}{2}[\mathbf{E}(x_k) +
  \mathbf{E}(x_{k+1})]\,.  
\end{equation}
The expectation value of $x_{k+1}$ is
calculated as above, using the distribution (\ref{eq:oddbias}), while
the distribution of $x_{k}$ is obtained by replacing $k$ with $k-1$
in (\ref{eq:oddbias}).  It turns out that $\mathbf{E}(\zeta^{(w)}) =
\mathbf{E}(\zeta^{(w-1)})$ when $w$ is even.  Thus
$\mathbf{E}(\zeta^{(2)}) = \mathbf{E}(\zeta^{(1)})$,
$\mathbf{E}(\zeta^{(4)}) = \mathbf{E}(\zeta^{(3)})$ and so on.  
Figure~\ref{fig:bias} plots $\mathbf{E}(\zeta^{(w)})$ for all values
of $w$ from $1$ to $200$.  
%

%%%%%%%%%%%%%%%%%%%%%%%%%%%%%%%%%%%%%%%%%%%%%%%%%%%%%%%%%%%%%%%%%%%%%%%%%%%%
\section{The number count outliers}
\label{app:outliers}

\begin{table}
\begin{tabular}{ccc}\hline
 Detector & Frequency (Hz) & Number count \\ 
\hline \hline 
 L1 & 210.36  & 268 \\
    & 234.50  & 224 \\
    & 281.35  & 218  \\  
    & 329.34  & 626\\
    & 335.62  & 219  \\       
\hline 
 H1 & 210.36  & 596 \\
    & 212.26  & 519 \\
    & 244.14  & 507 \\
    & 249.70  & 746 \\
    & 280.48  & 949 \\
    & 329.58  & 1510\\
    & 329.78  & 1227\\
    & 348.45  & 482 \\
    & 350.60  & 1423\\
\hline 
 H2 & 202.18  & 402  \\
    & 203.23  & 395  \\
    & 210.36  & 443 \\
    & 298.81  & 394 \\
    & 329.69  & 867 \\
    & 387.05  & 400 \\
    & 389.40  & 391 \\
\hline
\end{tabular}
\caption{List of outliers present in the Hough maps after applying
the known instrumental frequency veto, for a false alarm rate
of $10^{-13}$. This corresponds  to a threshold in the number count of 
216 for L1, 480 for H1 and 390 for H2. For each outlier 
we quote the central frequency and the maximum number count.
The triple coincidence at $210.36\,$Hz is a harmonic of the $70.12\,$Hz spectral
disturbance described in the text.} \label{tab:S2lines} 
\end{table}

This appendix  contains a discussion about the %unresolved
outliers in the Hough number counts that  are strongly suspected to be
instrumental artifacts but  that we were not able to definitely
identify as such. 

After applying the frequency veto described in Sec.
\ref{subsec:numberCounts},
we focus our attention on those candidates with a false alarm rate $\alpha$
(for a single detector) smaller than $10^{-13}$.
This corresponds to a threshold on the number count of 216 for L1, 480 for H1
and 390 for H2. Since the total number of templates analyzed in this 
$200\,$Hz search
band is roughly $10^{12}$, the probability of getting one candidate
above that threshold over the full search is approximately $10\%$ in
each detector. All the candidates that satisfy such condition tend to
cluster around a few frequencies that are listed in
Table~\ref{tab:S2lines}. These are the so-called  \emph{outliers} that
were present in Fig. \ref{fig:S2L1ncMax}, \ref{fig:S2H1ncMax} and
\ref{fig:S2H2ncMax}. 
%

%The fact that these outliers have such a small false alarm makes them
%unlikely to be noise, unless some spectral disturbance (line noise) mimics
%the time-frequency evolution of a pulsar for a certain location in the sky.
If these outliers are due to gravitational signals they should
show up in the different detectors. By performing a simple
coincidence analysis in frequency, the only candidate that remains, at this
false alarm level, is the one at $210.36\,$Hz.
The reader should notice that $210.36\,$Hz corresponds to $3\times 70.12\,$Hz.
In the H1 data, there are also excess of number counts at $280.480\,$Hz and 
$350.600\,$Hz, corresponding to $4\times 70.12$ and  $5\times 70.12$ respectively.
These $70.120\,$Hz multiples together with the $244.14\,$Hz line
were detected in association with a VME (VERSA module Eurocard) 
controller hardware used 
during S2. However, since the data acquisition system
architecture has changed since S2 the coupling mechanism
cannot be proven.

%arise from magnetic coupling of
%nearby computer monitor raster signals to actuation magnets on the test 
%mass mirrors.
%Correlations are observed between local magnetometer signals and the 
%gravitational wave data channel at the 70.120 Hz harmonics.
%{\color{red} More evidence for this connection, such as a smoking-gun 70.12 Hz
%emission from a monitor would be most welcome here!}.) 

\begin{figure}
  \begin{center}
  \includegraphics[height=9cm]{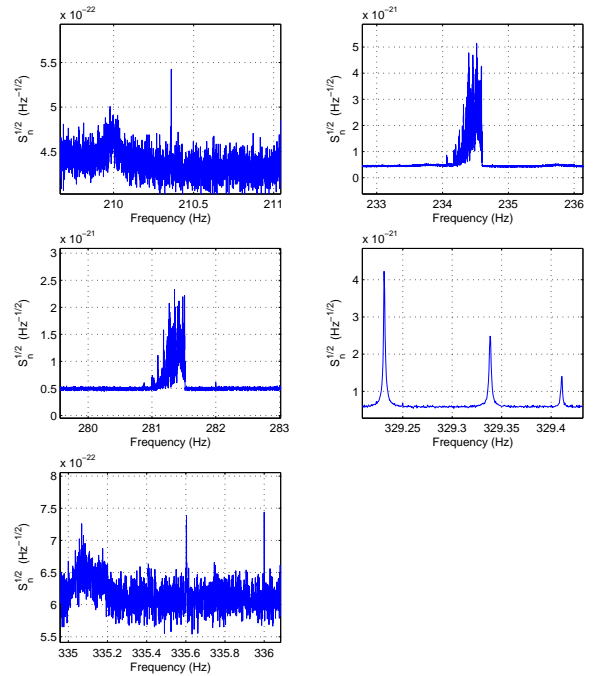}
  \caption{The square root of the average value of $S_n$ using the entire S2
  L1 data set analyzed. The four graphs correspond to the frequencies 
  where outliers were present in the Hough maps after applying the known
  instrumental veto. They correspond to $210.36\,$Hz, $234.50\,$Hz, $281.35\,$Hz, 
  $329.34\,$Hz and $335.62\,$Hz.}
  \label{fig:L1s2lines}
  \end{center}
\end{figure}

\begin{figure}
  \begin{center}
  \includegraphics[height=9.5cm]{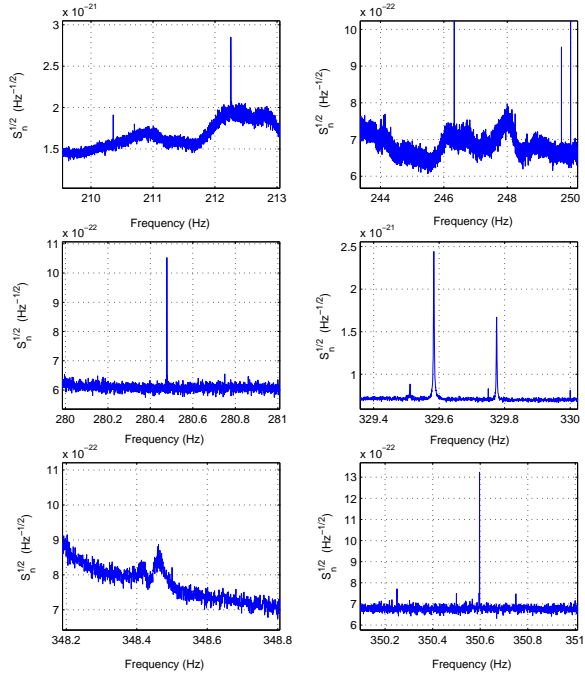}
  \caption{The square root of the average value of $S_n$ using the entire S2
  H1 data set analyzed. The graphs correspond to zooms in the frequencies 
  where outliers were present in the Hough maps. 
  They correspond to $210.36\,$Hz, $212.26\,$Hz,  
  $244.14\,$Hz, $249.70\,$Hz, $280.48\,$Hz, $329.58\,$Hz, $329.78\,$Hz,
  $348.45\,$Hz and $350.60\,$Hz.  }
  \label{fig:H1s2lines}
  \end{center}
\end{figure}

\begin{figure}
  \begin{center}
  \includegraphics[height=9cm]{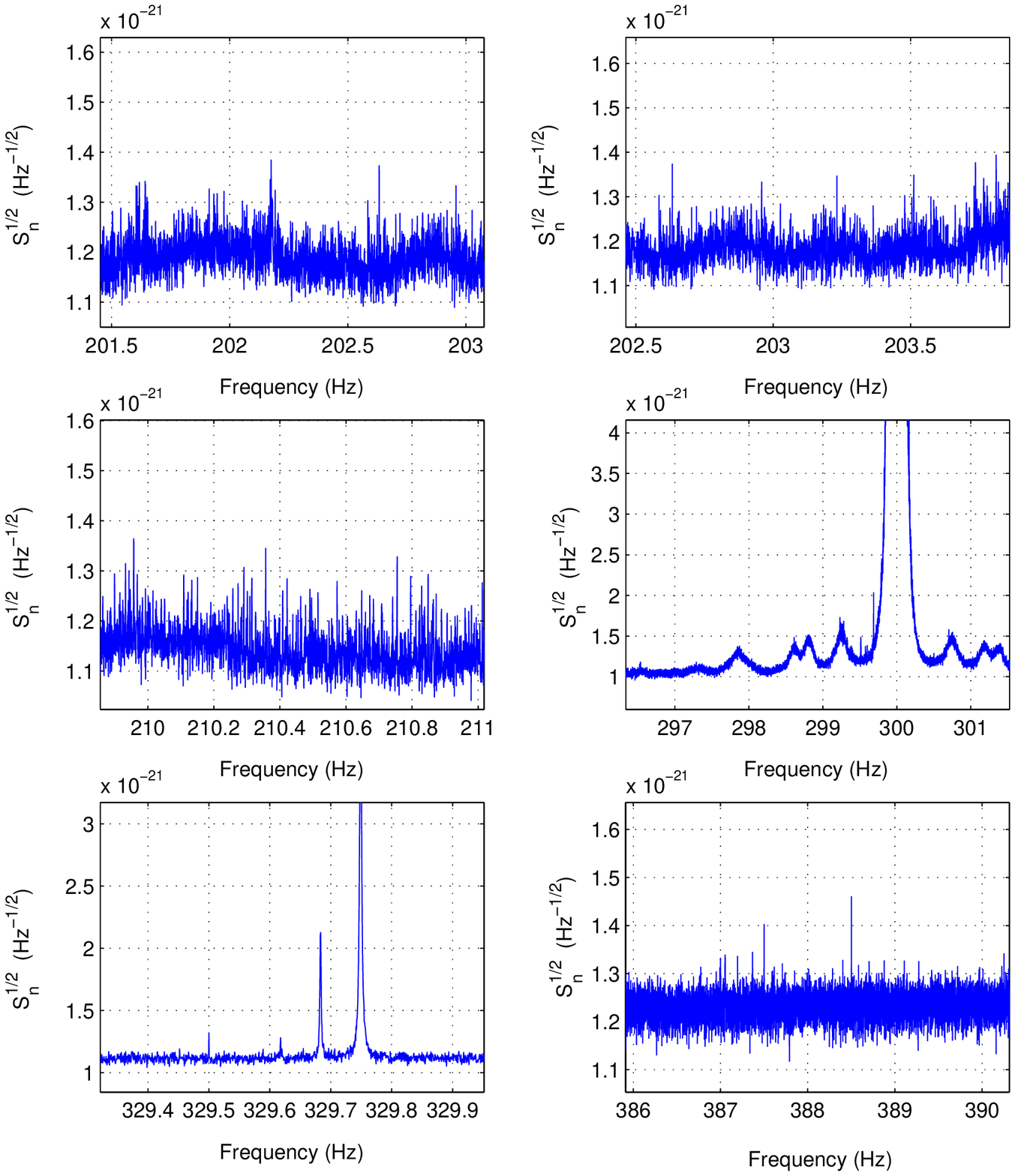}
  \caption{The square root of the average value of $S_n$ using the entire S2
  H2 data set analyzed. The graphs correspond to zooms in the frequencies 
  where outliers were present in the Hough maps after applying the known
  instrumental veto. Those correspond to $202.18\,$Hz, $203.23\,$Hz, $210.36\,$Hz, 
  $298.81\,$Hz, $329.69\,$Hz, $387.05\,$Hz and $389.40\,$Hz.}
  \label{fig:H2s2lines}
  \end{center}
\end{figure}

%We have estimated $S_n$ by averaging periodograms over the entire run.
%As shown in
Figs. \ref{fig:L1s2lines}, \ref{fig:H1s2lines} and \ref{fig:H2s2lines},
show how the 
the outliers listed in Table~\ref{tab:S2lines}
stand well above the background noise spectrum level $S_n$,
when this is estimated from the entire run.
We believe they all arise from instrumental or environmental
artifacts. However we are not able to determine in a 
conclusive manner their physical cause.
% although their nature is not completely understood.
For this reason, the potential contaminated frequency bands
have not been vetoed when setting the upper limits.

% However we are not able to determine in a 
%conclusive manner their physical cause.

%The strong lines around 329 Hz are also under investigation, 
%but despite their height well above the background noise level 
%no correlations have been found so far for L1 and H1 
%with the environmental channels, while for H2 we have seen a strong correlation
%with the signals from some microphones. {\color{red} More evidence would be
%welcome here too!}

\end{appendix}

%%%%%%%%%%%%%%%%%%%%%%%%%%%%%%%%%%%%%%%%%%%%%%%%%%%%%%%%%%%%%%%%%%%%%%
% references

%

\end{document}